\documentclass[manuscript]{aastex}

\usepackage[document]{ragged2e}
\usepackage{apjfonts}

\begin{document}

\title{Fluorescent H$_{2}$ Emission Lines from the Reflection Nebula NGC~7023 observed with IGRINS}

\author{
Huynh Anh N. Le$^{1,2}$, Soojong Pak$^1$\altaffilmark{,*}, Kyle Kaplan$^3$, Gregory Mace$^3$, Sungho Lee$^4$, Michael Pavel$^3$, Ueejeong Jeong$^4$, Heeyoung Oh$^{2,4}$, Hye-In Lee$^1$, Moo-Young Chun$^4$, In-Soo Yuk$^4$, Tae-Soo Pyo$^{5,6}$, Narae Hwang$^4$, Kang-Min Kim$^4$, Chan Park$^4$, Jae Sok Oh$^4$, Young Sam Yu$^4$, Byeong-Gon Park$^4$, Young Chol Minh$^4$, and Daniel T. Jaffe$^3$
}

\altaffiltext{1}{School of Space Research, Kyung Hee University, 1732 Deogyeong-daero, Giheung-gu, Yongin-si, Gyeonggi-do 17104, Korea; huynhanh7@khu.ac.kr}
\altaffiltext{2}{Department of Physics and Astronomy, Seoul National University, 1 Gwanak-ro, Gwanak-gu, Seoul 08826, Korea}
\altaffiltext{3}{Department of Astronomy, University of Texas at Austin, Austin, TX 78712, USA}
\altaffiltext{4}{Korea Astronomy and Space Science Institute, Daejeon 34055, Korea}
\altaffiltext{5}{Subaru Telescope, National Astronomical Observatory of Japan, National Institutes of Natural Sciences (NINS), 650 North Aohoku Place, Hilo, HI 96720,  USA}
\altaffiltext{6}{School of Mathematical and Physical Science, SOKENDAI (The Graduate University for Advanced Studies), Hayama, Kanagawa 240-0193, Japan}

\altaffiltext{*}{corresponding author; soojong@khu.ac.kr}

\begin{abstract}
\justify

We have analyzed the temperature, velocity and density of H$_{2}$ gas in NGC~7023 with a high-resolution near-infrared spectrum of the northwestern filament of the reflection nebula. By observing NGC~7023 in the $H$ and $K$ bands at $R$ $\simeq$ 45,000 with the Immersion GRating INfrared Spectrograph (IGRINS), we detected 68 H$_{2}$ emission lines within the 1$\arcsec$ $\times$ 15$\arcsec$ slit. The diagnostic ratios of 2-1 S(1)/1-0 S(1) is 0.41$-$0.56. In addition, the estimated ortho-to-para ratios (OPR) is 1.63$-$1.82, indicating that the H$_{2}$ emission transitions in the observed region arises mostly from gas excited by UV fluorescence. Gradients in the temperature, velocity, and OPR within the observed area imply motion of the photodissociation region (PDR) relative to the molecular cloud. In addition, we derive the column density of H$_{2}$ from the observed emission lines and compare these results with PDR models in the literature covering a range of densities and incident UV field intensities. The notable difference between PDR model predictions and the observed data, in high rotational $J$ levels of $\nu$ $=$ 1, is that the predicted formation temperature for newly-formed H$_{2}$ should be lower than that of the model predictions. To investigate the density distribution, we combine pixels in 1$\arcsec$ $\times$ 1$\arcsec$ areas and derive the density distribution at the 0.002 pc scale. The derived gradient of density suggests that NGC~7023 has a clumpy structure, including a high clump density of $\sim$10$^{5}$ cm$^{-3}$ with a size smaller than $\sim$5 $\times$ 10$^{-3}$ pc embedded in lower density regions of 10$^{3}$$-$10$^{4}$ cm$^{-3}$.

\end{abstract}


\section{Introduction}\label{intro}
\justify

Molecular hydrogen H$_{2}$ is a major component of the interstellar medium (ISM). Rovibrational H$_{2}$ emission lines arise either in shock-heated regions or in photodissociation regions (PDRs). In PDRs, the far-UV (FUV) photons illuminate the transition layer between the ionized gas and the surface of the molecular cloud (\citealp{Tielens85a}; \citealp{Tielens85b}; \citealp{Black87}; \citealp{Sternberg89}; \citealp{Burton89}; \citealp{Draine96}; \citealp{Luhman97}; \citealp{Hollenbach99}). Molecular hydrogen emission has been observed in reflection nebulae (e.g., NGC~2023: \citealp{Gatley87}; \citealp{Sellgren86}; \citealp{Hasegawa87}; \citealp{Burton98}; \citealp{Martini99}; \citealp{McCartney99}; \citealp{Habart04,Habart11}; \citealp{Sheffer11}; \citealp{Fleming10}), the Orion nebula (\citealp{Hayashi85}; \citealp{Luhmank96,Luhmank98}; \citealp{Bertoldi99}; \citealp{Rosenthal00}; \citealp{Allers05}; \citealp{Habart04}), M17 (\citealp{Chrysostomou92,Chrysostomou93}; \citealp{Sheffer13}), and in planetary nebulae (e.g., Hubble 12: \citealp{Ramsay93}; \citealp{Hora96}; \citealp{Chrysostomou98}; \citealp{Marquez15}).


In general, H$_{2}$ emission lines at $\nu$ = 0 and low rotational $J$ levels are good indicators of the gas temperature in the layer where they arise, while H$_{2}$ lines arising from higher excitation energy states are pumped by UV photons and are sensitive to the radiation field, gas temperature, and gas density. Theoretical PDR models predict the intensities and line ratios of H$_{2}$ in PDRs well (\citealp{Sternberg89}; \citealp{Draine96}; \citealp{Shaw05}; \citealp{Shaw09}). However, observations of rotational H$_{2}$ emission lines by ISO (the Infrared Space Observatory, \citealp{Kessler96}) indicate that the gas temperature derived from first rotational levels of H$_{2}$ is higher than that predicted in the PDR models (\citealp{Timmermann96}; \citealp{Fuente99}; \citealp{Bertoldi99}; \citealp{Draine99}; \citealp{Thi00}). Theoretical models have difficulty in explaining the existence of much warmer H$_{2}$ rotational emission zones. Modification of the heating and cooling processes, or H$_{2}$ formation rates, have been proposed to explain these discrepancies (\citealp{Habart04}; \citealp{Draine99}; \citealp{Allers05}). \citet{Weingartner99} suggest that radiation forces on dust grains can enhance the dust-to-gas ratio, thus increasing the gas heating rate.


Discrepancies between the observed intensities of fluorescent H$_{2}$ emission from highly excited energy levels, or high rotational states, and those predicted by PDR models are a source of debate. \citet{Burton98} indicated that the observed intensities of high $\nu$ lines are weaker than model predictions. Yet, \citet{Bertoldi00} pointed out that the observed intensities of H$_{2}$ emission lines at very high energy states ($\nu$ = 9 and 12) were stronger than the values predicted by their PDR model. In addition, H$_{2}$ lines arising from high-$\nu$ or high-$J$ states are sensitive to UV field intensity and gas density. The difference between model results and observations may be explained by the effects of collisional de-excitation of the UV excited H$_{2}$, the formation H$_{2}$, or changing ortho-to-para ratios \citep{Burton98}. \citet{Burton92b} suggested that H$_{2}$ formation has a significant effect on the $\nu$ = 4 level from observations of NGC~2023 but the level distribution of newly-formed H$_{2}$ is unclear (\citealp{Draine96}, \citealp{Bertoldi99}). To date, most of the observations of fluorescently excited H$_{2}$ emission lines have been taken at low spatial and spectral resolution. Those spectra do not have enough resolution to robustly test H$_{2}$ line intensities. At high spatial and spectral resolution, H$_{2}$ emission lines resolve the physical structure of PDRs and test the predictions and assumptions of fluorescent PDR models.

The reflection nebula NGC~7023 is illuminated by the Herbig B3Ve$-$B5 star HD~200775 (\citealp{Witt06}; \citealp{Alecian08,Alecian13}), with an effective temperature of 17,000 K \citep{Baschek82}, at a distance of $430_{-90}^{+160}$ pc \citep{van97}. The proximity of this object to us and the known properties of the exciting star make it one of the best objects for studying PDRs. Observations show that this nebula hosts different gas density structures, including dense clumps (n $\sim$ 10$^{6}$ cm$^{-3}$) embedded in lower density gas (n $=$ 10$^{4}$$-$10$^{5}$ cm$^{-3}$) (\citealp{Chokshi88}; \citealp{Sellgren92}; \citealp{Fuente96}; \citealp{Lemaire96}; \citealp{Martini97}; \citealp{Martini99}; \citealp{Fuente00}; \citealp{Takami00}; \citealp{An03}; \citealp{Fleming10}; \citealp{Habart11}; \citealp{Kohler14}; and \citealp{Pilleri12,Pilleri15}). At the wall of the cavity, PDR emission arises $\sim$42$\arcsec$ northwest, $\sim$55$\arcsec$ southwest, and $\sim$155$\arcsec$ east of the central illuminating star where the FUV field intensities of $G$ = 2600, 1500, and 250, respectively, in units of {\it G$_{0}$} $=$ 1.6 $\times$ 10$^{-3}$ erg cm$^{-2}$ s$^{-1}$ \citep{Pilleri12}.

NGC~7023 has been observed in the near-infrared (near-IR) by many authors (e.g., \citealp{Sellgren86}; \citealp{Sellgren92}; \citealp{Lemaire96}; \citealp{Lemaire99}; \citealp{Martini97}; \citealp{Martini99}). In particular, low spectral resolution observations of H$_{2}$ emission (\citealp{Martini97,Martini99}) found high density clumps in the PDR regions of NGC~7023. In addition, \citet{Lemaire96} found small scale structure, less than 0.004 pc, based on high spatial resolution images of $\nu$ = 1-0 S(1) and S(2) rovibrational lines.


The Immersion Grating Infrared Spectrograph (IGRINS) covers the whole $H$ and $K$ bands ($1.4$$-$$2.5$ $\mu$m) in a single exposure with a resolving power of $R$ $\simeq$ 45,000 (\citealp{Yuk10}; \citealp{Park14}). Since its commissioning in 2014, many interesting results in the ISM have been explored by (\citealp{Lee15}; \citealp{Sterling16}; \citealp{Afsar16}; \citealp{Oh16a}; \citealp{Oh16b}; \citealp{Lee16}; \citealp{Kaplan17}).

With our IGRINS observations we can isolate the physical structure and properties in the observed region with high spectral and spatial resolution, and study many H$_{2}$ transitions simultaneously with exact spatial registration of the observations at different wavelengths. In this paper, we present detections of many H$_{2}$ lines from the northwestern (NW) filament of NGC~7023 and derive the distribution of temperature, column density, and ortho-to-para ratio of H$_{2}$ in the observed area. We also test the reliability of PDR models at the \ion{H}{2} and H$_{2}$ interface in NGC~7023.



\section{Observations and Data Reduction}\label{obs}

\subsection{Observations}

\begin{figure*}[p]
\figurenum{1}
\centering
\includegraphics[width=50ex,height=50ex, keepaspectratio]{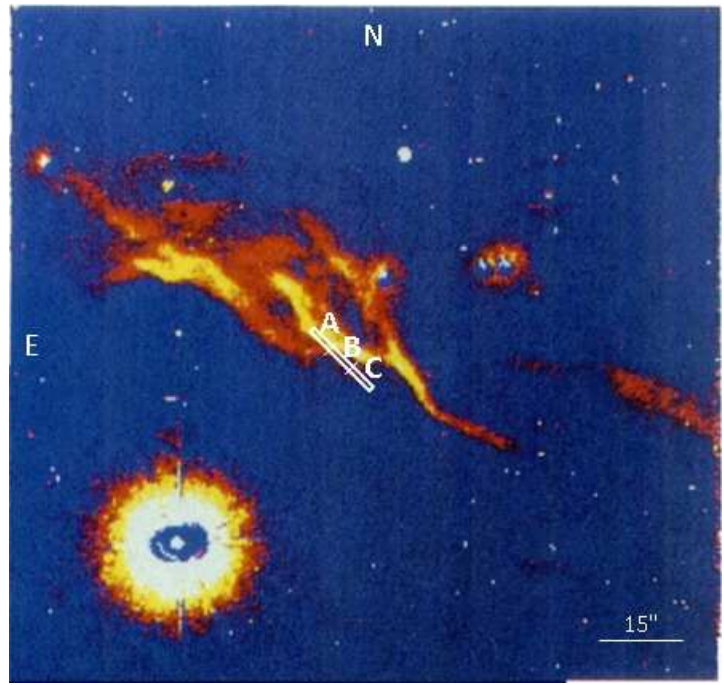}
\caption{Image of the H$_{2}$ emission at 2.121 $\mu$m from the northwestern filament of NGC~7023 recorded in \citet{Lemaire96}. In the image, the central illuminating star, HD~200775, lies at $\alpha$(2000) = 21$^{h}$01$^{m}$36$^{s}$.9 and $\delta$(2000) = 68$^{\circ}$09$\arcmin$47$\farcs$8. The IGRINS slit center position is located at 29$\farcs$0 west, 32$\farcs$7 north of the star with PA=45$^{\circ}$. The position of regions A, B, and C are shown as text in the image.}
\label{slitview}
\end{figure*}

We observed NGC~7023 on 2014 July 12 during IGRINS commissioning on the 2.7 m Harlan J. Smith Telescope at the McDonald Observatory. The slit position angle was 45$^{\circ}$ (east from north) for this observation and the slit size was fixed at 1$\arcsec$ $\times$ 15$\arcsec$. The slit center position (SC) in the nebula was 29$\farcs$0 west, 32$\farcs$7 north of the central illuminating star, HD~200775 which lies at $\alpha$(2000) $=$ 21$^{h}$01$^{m}$36.9$^{s}$ and $\delta$(2000) $=$ 68$^{\circ}$09$\arcmin$47$\farcs$8. We use the ``Nod-off-slit'' mode with an ``off'' position 120$\arcsec$ to the north from the target. The total on-source exposure-time was 1200~s. Figure \ref{slitview} shows the position of the slit on an H$_{2}$ 2.121 $\mu$m emission map \citep{Lemaire96}. Based on the full width at half maximum (FWHM) of the bright star in the slit-view camera image of NGC~7023, the seeing during the observation was $\sim$1$\farcs$0 in the $K$ band.

\subsection{Data Reduction}

Data reduction was performed using the IGRINS data reduction pipeline package\footnote{See also the IGRINS pipeline package at https://github.com/igrins/plp/tree/v2.0.} \citep{Leejj15}. The standard data reduction processes including bad-pixel and cosmic ray correction, flat fielding, sky subtraction, spectral extraction, and wavelength calibration are described in \citet{Sim14}. We corrected bad-pixels and cosmic rays for all of the target and sky frames. A combined master flat image was made to correct variations of pixel-to-pixel in the detector and to do aperture extraction maps for tracing the orders in the echellogram. Thorium-argon (ThAr) emission lines were used to do distortion correction and wavelength calibration. Sky background was corrected by subtracting the sky frame from the target frame. Telluric absorption correction and absolute flux calibration was achieved by dividing by the spectrum of the A0V standard star HD~155379 observed at a similar airmass to NGC~7023.

\subsection{H$_{2}$ Lines Collection and OH-Residuals}

\begin{figure*}[p]
\figurenum{2a}
\centering
\includegraphics[width=100ex, angle=90]{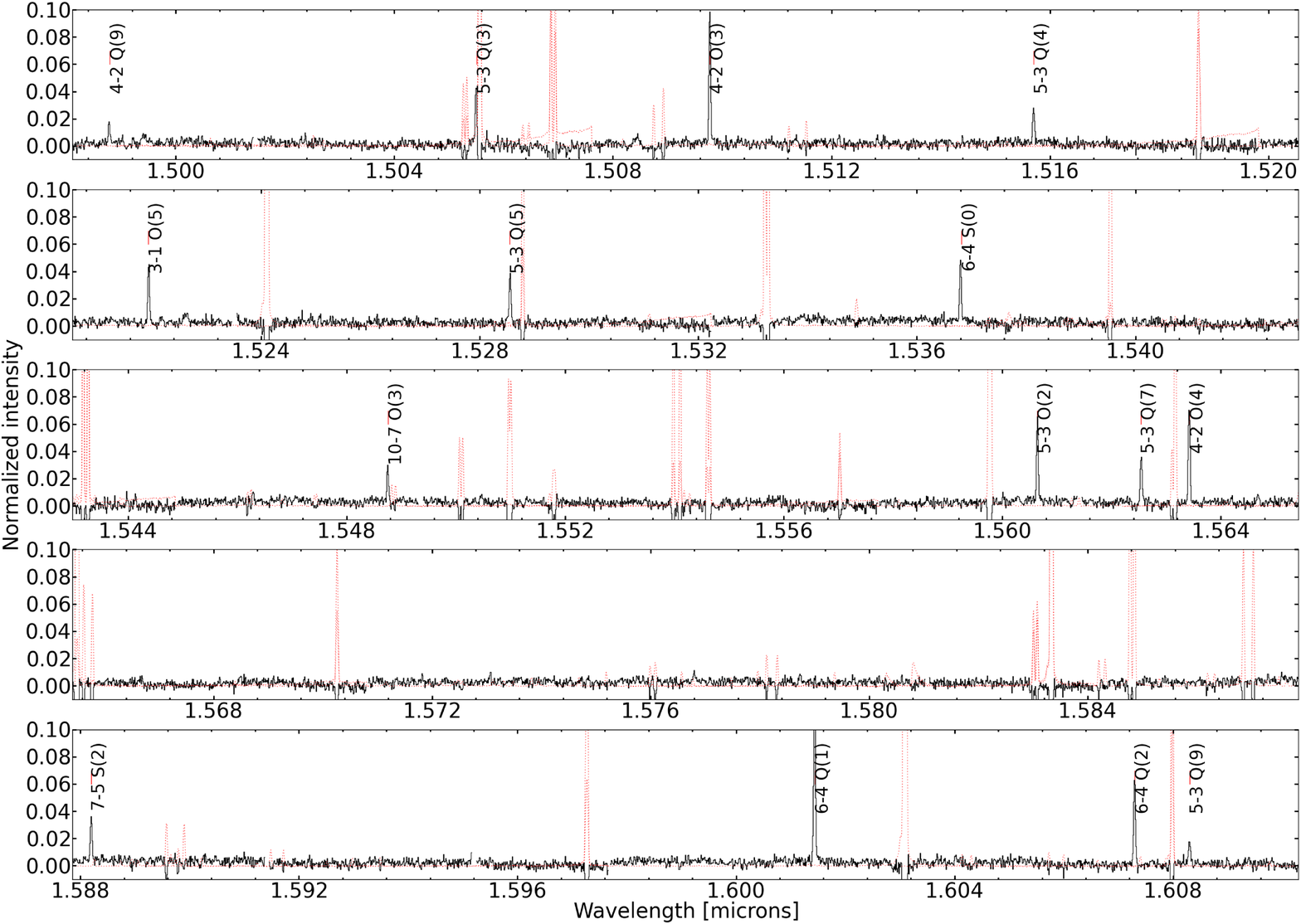}
\caption{\label{hband1} Part of the $H$ band spectrum in 1.497$-$1.610 $\mu$m bands from the NW filament of NGC~7023. The intensity has been normalized by the peak of the 1-0 S(1) line. The normalized intensity of H$_{2}$ 1-0 S(1) 2.1218 $\mu$m is one. We marked the identified lines with their names in the figure. The dash-red lines display OH emission lines, observed at ``off'' position 120$\arcsec$ to the north from the target.}
\end{figure*}

\begin{figure}[p]
\figurenum{2b}
\centering
\includegraphics[width=100ex, angle=90]{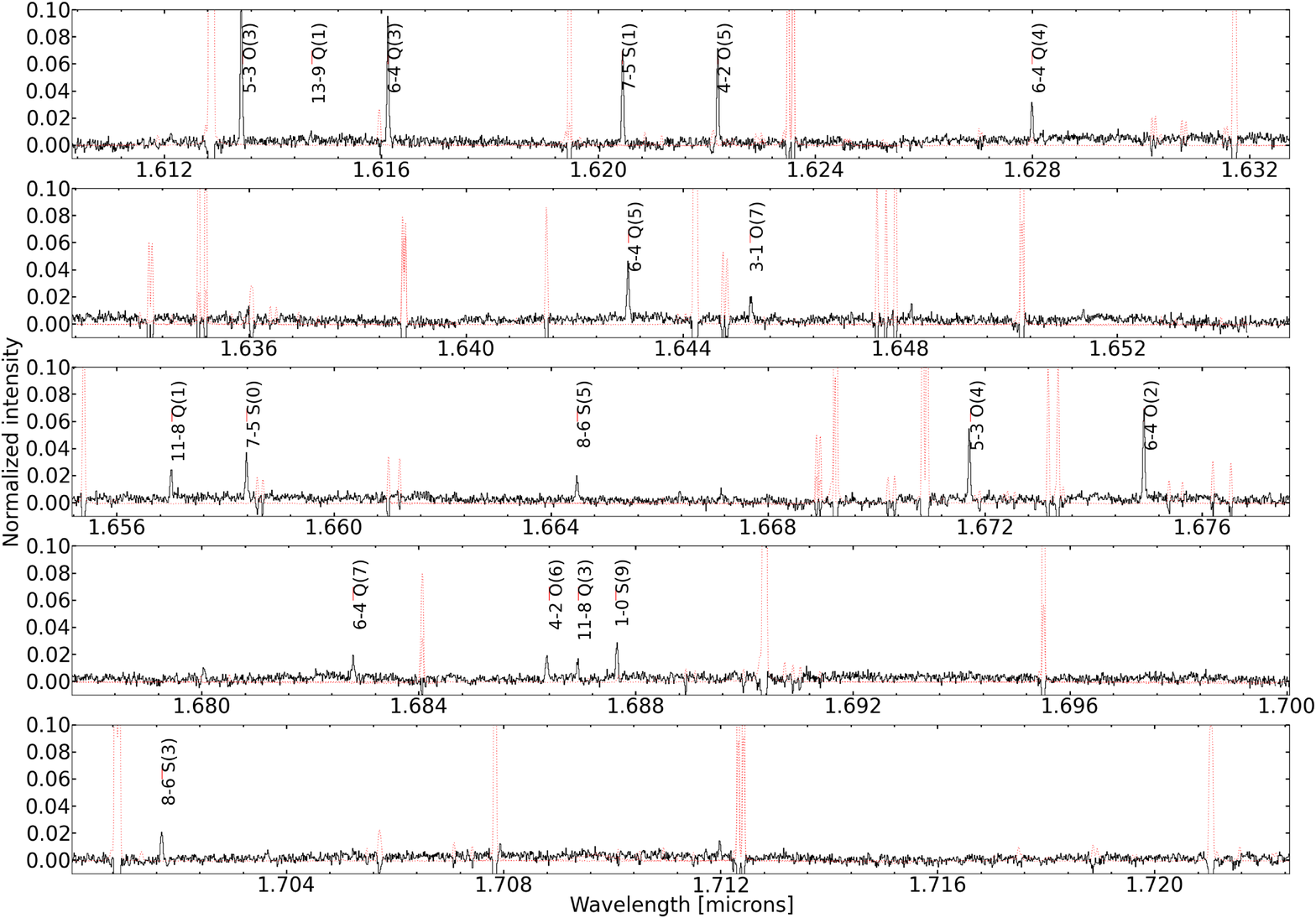}
\caption{\label{hband2} Same as Figure 2a, except for the wavelength range is in 1.610$-$1.722 $\mu$m bands.}
\end{figure}

\begin{figure}[p]
\figurenum{2c}
\centering
\includegraphics[width=100ex, angle=90]{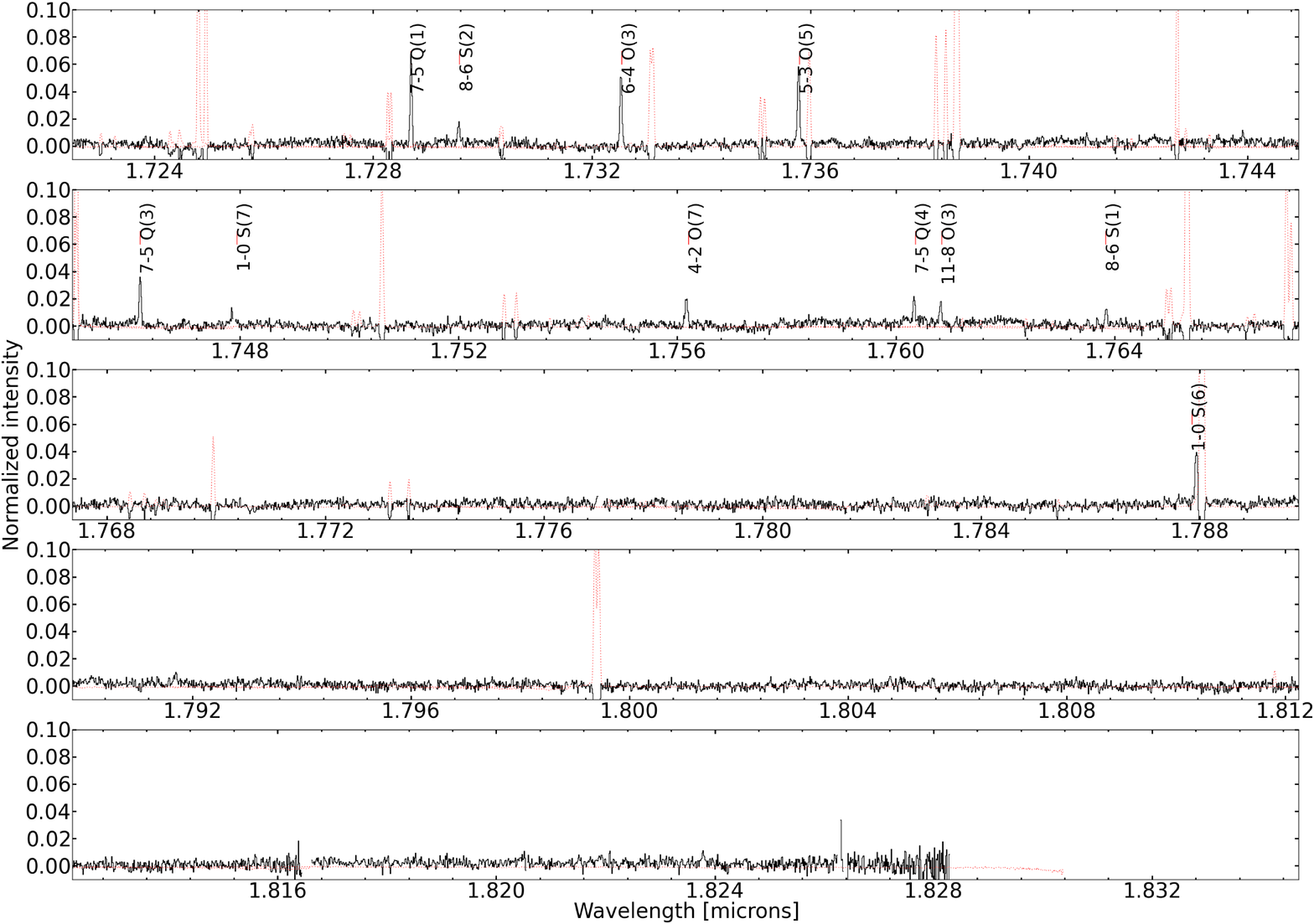}
\caption{\label{hband3} Same as Figure 2a, except for the wavelength range is in 1.722$-$1.832 $\mu$m bands.}
\end{figure}

\begin{figure}[p]
\centering
\figurenum{2d}
\includegraphics[width=100ex, angle=90]{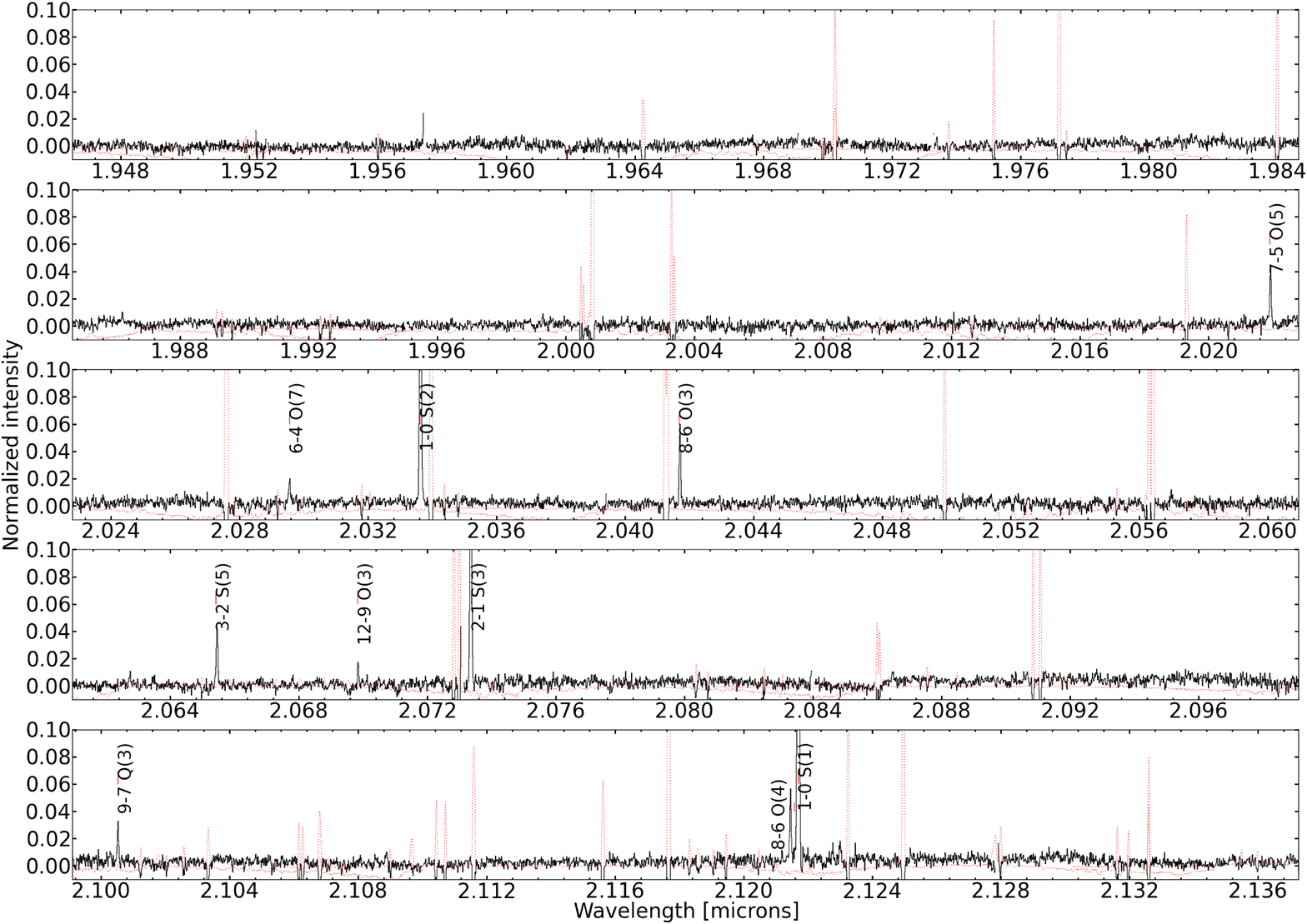}
\caption{\label{kband1} Same as Figure 2a, except for the wavelength range is in 1.946$-$2.137 $\mu$m bands.}
\end{figure}

\begin{figure}[p]
\figurenum{2e}
\centering
\includegraphics[width=100ex, angle=90]{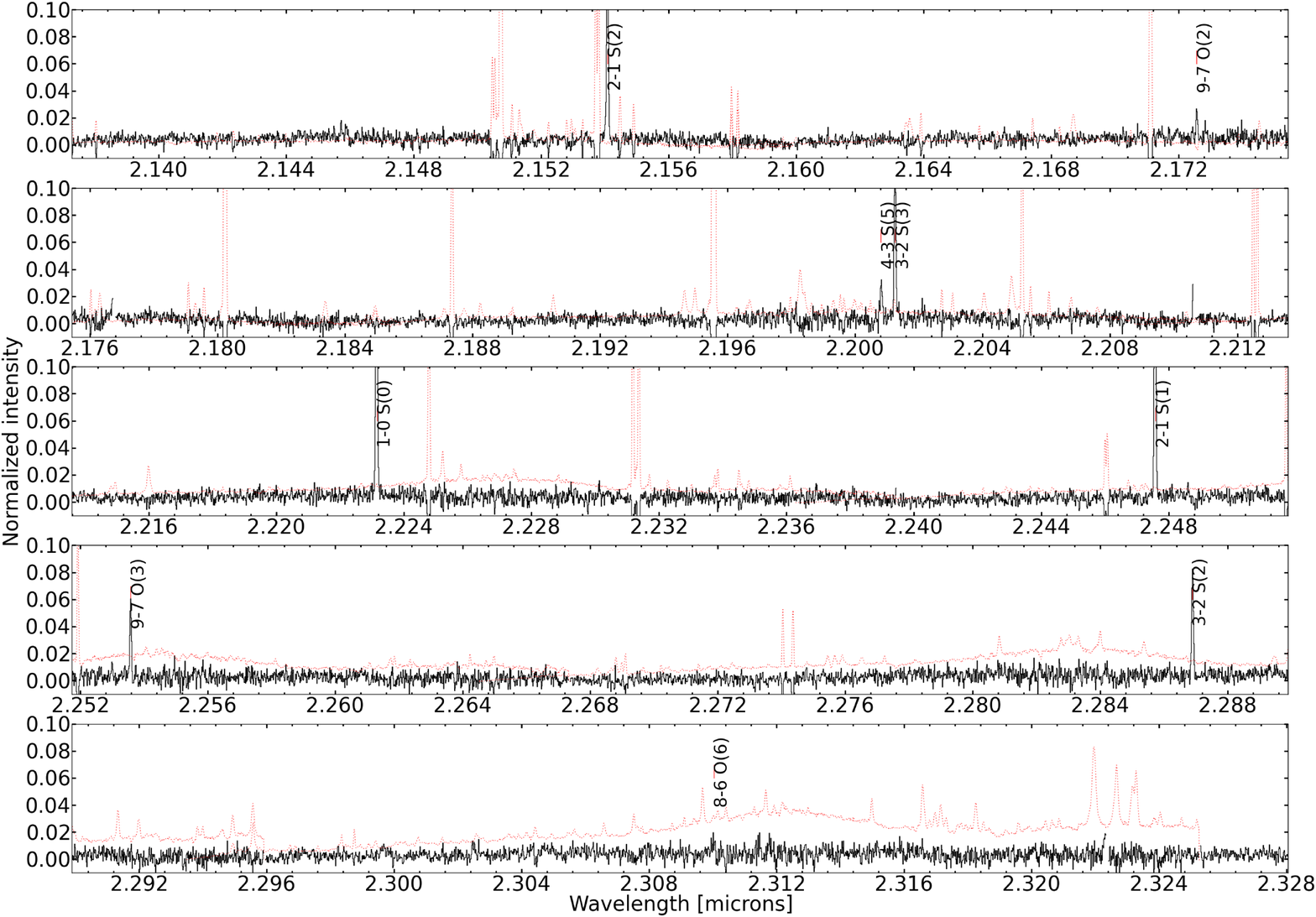}
\caption{\label{kband2} Same as Figure 2a, except for the wavelength range is in 2.137$-$2.328 $\mu$m bands.}
\end{figure}

\begin{figure}[p]
\figurenum{2f}
\centering
\includegraphics[width=100ex, angle=90]{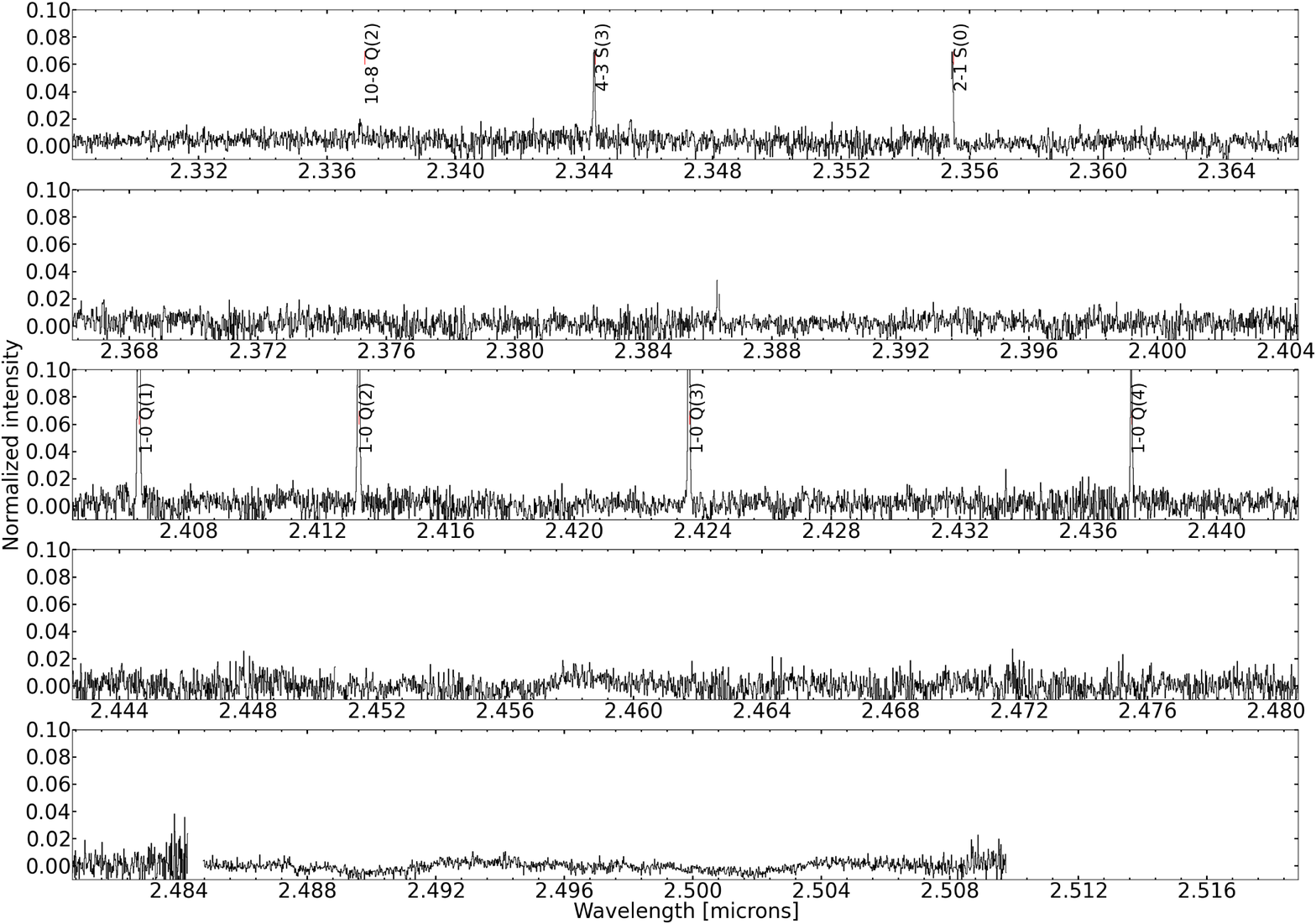}
\caption{\label{kband3} Same as Figure 2a, except for the wavelength range is in 2.328$-$2.512 $\mu$m bands.}
\end{figure}

\begin{figure*}[p]
\figurenum{3}
\centering
\includegraphics[width=90ex,height=90ex, keepaspectratio]{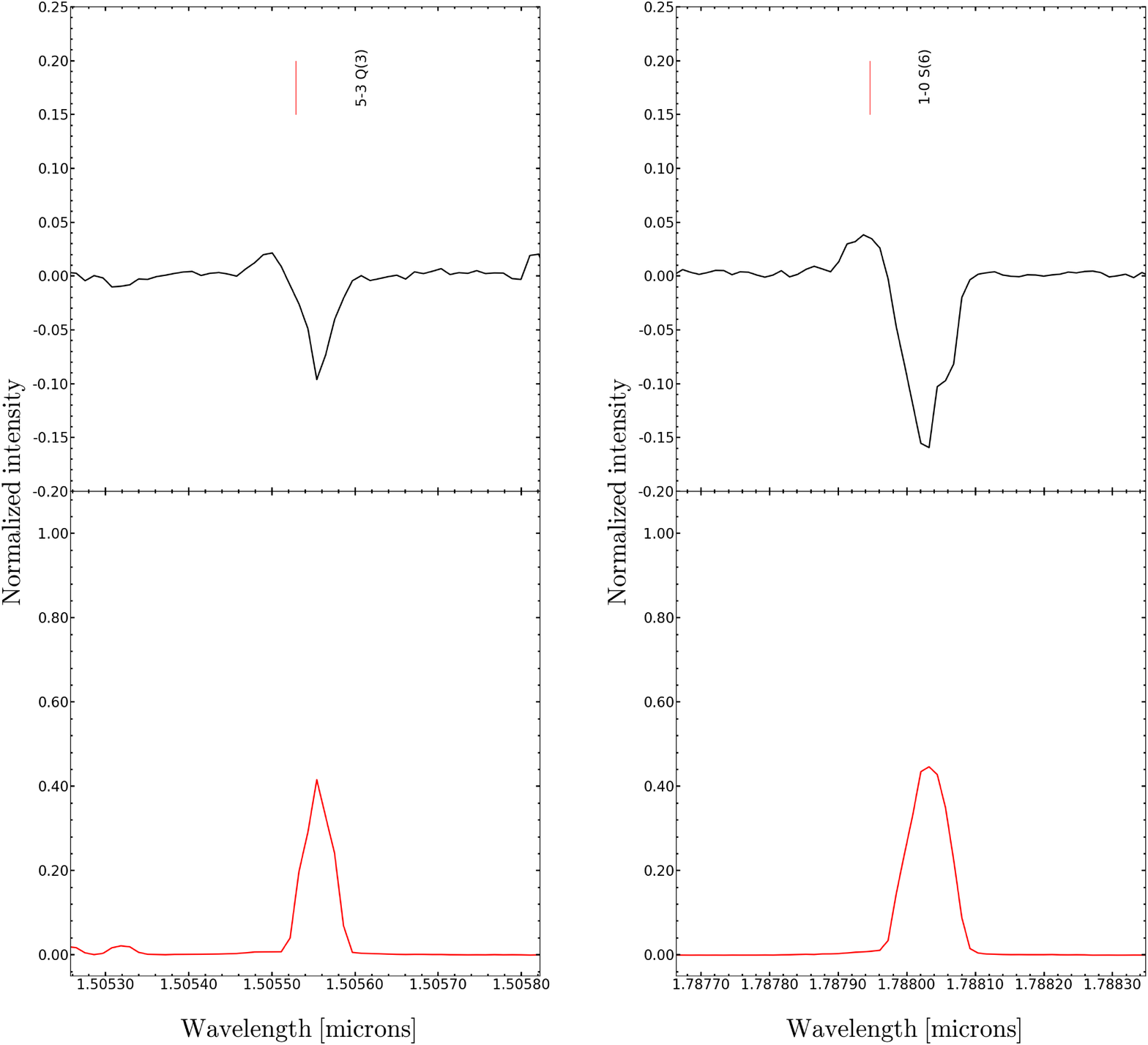}
\caption{Observed H$_{2}$ 5-0 O(3) and 1-0 S(6) line profiles (black-solid lines). The intensity has been normalized by the peak of the 1-0 S(1) line. The OH emission lines (red-solid lines) are shown in the bottom plots.}
\label{distube}
\end{figure*}

OH lines can not be completely removed from the target frame because telluric OH line fluxes vary over time. Residual OH emission lines may be misidentified as H$_{2}$ lines or disturbe the identified lines. Figures \ref{hband1}$-$\ref{kband3} show the whole spectrum from the NW filament of NGC~7023 with the intensity values normalized to the 1-0 S(1) intensity. All of the identified lines are marked in the figures. We used the list of wavelength of H$_{2}$ transitions in \citet{Draine96}, and detected 68 H$_{2}$ rovibrational emission lines. In addition, we also overplot the OH airglow emission lines in the off-position spectra in Figures \ref{hband1}$-$\ref{kband3} to check whether the H$_{2}$ lines are contaminated by OH-residuals or not. We carefully checked the measured wavelength of the detected lines by comparing with the published wavelength of the OH lines in $H$ and $K$ band spectra \citep{Rousselot00}. Then, we confirm that the H$_{2}$ line identifications are reliable. In the cases of 5-3 Q(3) and 1-0 S(6), we note that those lines are blended since they are strongly disturbed by OH-residuals after a subtraction of the OH emission lines. Figure \ref{distube} shows the observed 5-3 Q(3) and 1-0 S(6) line profiles.

\section{Analysis}\label{result}

\subsection{Flux Calibration}

We measured and calibrated the intensity using the methods in \citep{Le15}. We integrate the source signal within $\pm$12 km s$^{-1}$ of the line center, because the typical observed full width at half maximum (FWHM) of the detected H$_{2}$ lines is $\sim$ 8 km s$^{-1}$ (dominated by instrument line width). The noise was estimated by combining the source noise from the line flux itself with the detector read noise in quadrature.



The flux calibration is done by converting pixel values in ADU into specific intensities in physical units (e.g., W m$^{-2}$ arcsec$^{-2}$), for which we follow the methods in \citet{Lee06}. By calculating the flux calibration scaling factor $\Re_{\mathrm{ext}}$, the flux calibration is determined by,

\begin{equation}\label{eq:flux}
I(\lambda )= \Re_{\mathrm{ext}}(\lambda )  S(\lambda )
\end{equation}
where $I(\lambda$) is specific intensity in units of W~m$^{-2}$~$\mu$m$^{-1}$ arcsec$^{-2}$, and $S(\lambda$) is in ADU units. The scaling factor of flux calibration $\Re_{\mathrm{ext}}$ is calculated by,

\begin{equation}\label{eq:cflux_extend}
\Re_{\mathrm{ext}}(\lambda ) = \frac{F_{\mathrm{std}}(\lambda )}{S_{\mathrm{std}}(\lambda )}\frac{1}{\Omega}\frac{t_{\mathrm{std}}}{t_{\mathrm{obj}}}\frac{1}{W_{\mathrm{slit}}\tau_{\mathrm{slit}}}
\end{equation}
where $F_{\mathrm{std}}$ is the flux density of the standard star HD~155379 in W m$^{-2}$ $\mu$m$^{-1}$ with the K-magnitude of $m_{\mathrm{k}}$ = 6.52 mag, $S_{\mathrm{std}}$($\lambda$) is the integrated pixel value of the standard star in ADU units, $\Omega$ = 0$\farcs$27 $\times$ 0$\farcs$27 arcsec$^{2}$ is the field-of-view (FOV, 1 pixel = 0$\farcs$27), $t_{\mathrm{std}}$ = 480 s and $t_{\mathrm{obj}}$ = 1200 s are the exposure-time of the standard star and the object, respectively, $W_{\mathrm{slit}}$ = 3.66 pixels is the width of the slit. Note that there is a typo of the equation (8) in \citet{Lee06}. The slit obscuration factor $\tau_{\mathrm{slit}}$ should be

\begin{equation}\label{eq:tau}
\tau_{\mathrm{slit}} = \frac{1}{erf(\frac{Y}{\sqrt{2}\sigma })}
\end{equation}
where Y is half of the slit-width, Y = 0$\farcs$5, and $\sigma$ is the Gaussian width of the point spread function (PSF) along the slit-length. The measured $\sigma$ in our IGRINS spectrum is 0$\farcs$84.

%

\subsection{H$_{2}$ Spectra}\label{spectra}

\begin{figure*}[p]
\figurenum{4}
\centering
\includegraphics[width=70ex,height=70ex, keepaspectratio]{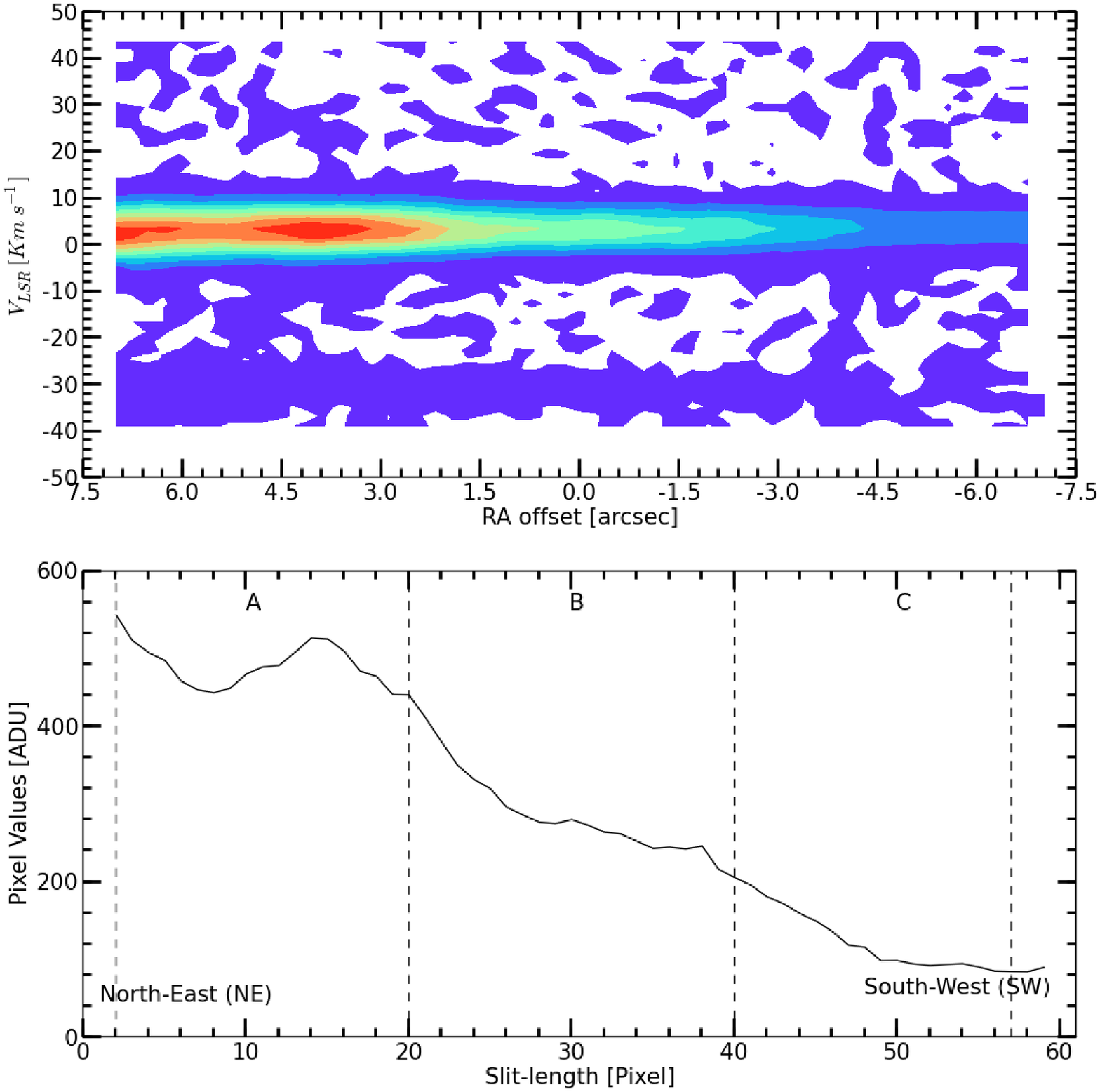}
\caption{Radial velocity diagram of the 1-0 S(1) line (top plot). In the diagram, 1$\arcsec$ = 3.66 pixels. The bottom plot shows the intensity profile at the peak of the 1-0 S(1) line. The black-dash lines show the separated regions, A, B, and C along the slit-length.}
\label{strip}
\end{figure*}

Figure \ref{strip} shows a position-radial velocity diagram of the H$_{2}$ 1-0 S(1) line. The peak intensity profile of the line is also displayed. We divided the observed region along the slit-length from north-east (NE) to south-west (SW) into regions of A (offset 7$\farcs$5$-$2$\farcs$5 to the NE from SC), B (offset 2$\farcs$5 to the NE and 2$\farcs$5 to the SW from SC), and C (offset 2$\farcs$5$-$7$\farcs$5 to the SW from SC).

We measured the intensity of 68 H$_{2}$ rovibrational emission lines arising in vibrational ladders from $\nu$ $=$ 1$-$13 and from rotational upper states as high as $J$ $=$ 11. For regions A, B and C we find that the upper states span an excitation temperature range of $E/k$ = $6100$$-$$50000$ K.

In order to correct the measured intensity values of H$_{2}$ emission lines for dust extinction, we applied a visual extinction of A$_{V}$ $=$ 2.2 mag, derived from R$_{V}$ $=$ 5 \citep{Mathis90} and E(B-V) $=$ 0.44 \citep{Witt80} as adopted by \citet{Martini97,Martini99}. We also used the IR extinction law in Equation \ref{eq:extinc} (\citealp{Cox00} and references therein) to correct for extinction across the entire IGRINS spectral range.

\begin{equation}\label{eq:extinc}
A(\lambda) = 0.412 A_{V} \lambda^{-1.75}
\end{equation}

Table \ref{h2line} shows the H$_{2}$ intensities listing only those lines with signal-to-noise greater than 5. Note that the real extinction values would be different from the adopted A$_{V}$ which we used to correct the intensity values of H$_{2}$ emission lines. This difference may affect our intensity measurements. We will discuss the effects of the adopted extinction value to our results in the discussion section.


\subsection{Dynamical Information from H$_{2}$ Lines} \label{h2shift}

In Table \ref{dynamic}, we present the line widths and velocities of the detected H$_{2}$ lines. Since we expect telluric absorption and OH airglow emission lines to contaminate our target spectra, we calculated the average velocity shift and line width of the detected H$_{2}$ emission lines and used those values as a reference for filtering the detected emission lines. We assume that all lines are from the same region and that reliable H$_{2}$ emission lines have velocities and line widths within the standard deviation of the detected emission lines, $\sigma_{V_{\mathrm{shift}}}$ and $\sigma_{V_{\mathrm{obs}}}$. The line shifts in wavelength locations have been corrected for the Earth and heliocentric velocities (8.63 km s$^{-1}$) and the local standard of rest velocity (LSR, 13.89 km s$^{-1}$), obtained by using the RVCORRECT task from {\it IRAF} \footnote{{\it IRAF} (Image Reduction and Analysis Facility) is distributed by the National Optical Astronomy Observatories (NOAO).}.

The radial velocities in $V_{\mathrm{LSR}}$ of the H$_{2}$ emission lines are 2-5 km s$^{-1}$. The average $V_{\mathrm{LSR}}$ of the H$_{2}$ lines for regions A, B, and C are 3.64 $\pm$ 0.16 km s$^{-1}$, 3.46 $\pm$ 0.11 km s$^{-1}$, and 3.39 $\pm$ 0.22 km s$^{-1}$, respectively. The average from all three regions is 3.50 $\pm$ 0.16 km s$^{-1}$, in agreement with the radial velocity of the 1-0 S(1) line given by \citet{Lemaire99} for NGC~7023 at the position which is $\sim$5$\arcsec$ offset from ours.

With the high spectral resolution of IGRINS, we were able to deconvolve the instrument profile from the H$_{2}$ emission lines. The measured FWHM was 7-9 km s$^{-1}$. Based on a comparison of the measured profiles of unresolved lines from our discharge ThAr arc calibration lamp, we find that a Gaussian is a good approximation to the instrumental profile. We therefore assume that the profile is Gaussian when deconvolving the spectra. We measured the line width of the ThAr emission lines from an internal IGRINS calibration lamp, observed from the same night observations, and we used them as the line width of the instrument profile. We estimated the FWHM of the H$_{2}$ lines by subtracting in quadrature the instrumental profile determined from the same-night calibration measurements of a ThAr emission-line spectrum.

\begin{figure*}[p]
\figurenum{5}
\centering
\includegraphics[width=70ex,height=70ex, keepaspectratio]{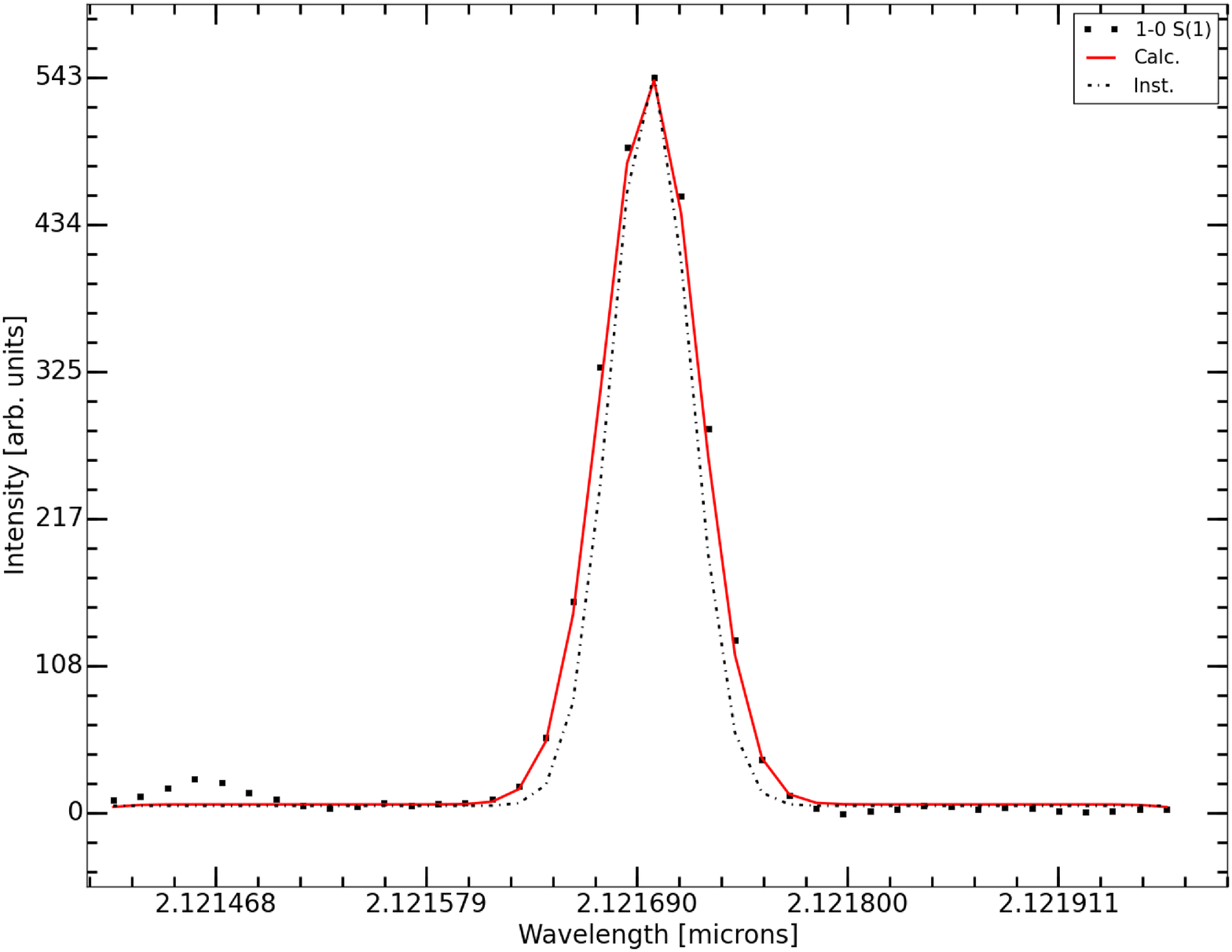}
\caption{Observed H$_{2}$ 1-0 S(1) line profile from region A in black-square symbols. The black-dotted points present the instrument profile. The solid-red line is from the convolution of the instrument profile and the derived intrinsic line width.}
\label{h2dynamic}
\end{figure*}

Figure \ref{h2dynamic} shows one example of the H$_{2}$ 1-0 S(1) line, the instrument profile, and the Gaussian fit by deconvolving the instrument profile from the 1-0 S(1) line in region A. The derived intrinsic line width is 2$-$6 km s$^{-1}$. We found that the average line widths of regions A, B and C were 4.63 $\pm$ 0.34 km s$^{-1}$, 3.29 $\pm$ 0.26 km s$^{-1}$, and 3.20 $\pm$ 0.27 km s$^{-1}$, respectively. Those values correspond to the upper limits of the kinematic temperatures of $\sim$930 K, $\sim$470 K, and $\sim$440 K, if all gas motions are thermal. Our derived line widths are consistent with the published result from \citet{Lemaire99} for the 1-0 S(1) line, 3.4 $\pm$ 0.8 km s$^{-1}$.

\section{Results}\label{disscus}

\subsection{H$_{2}$ Line Ratios}

The ratio of the line intensity, 2-1 S(1)/1-0 S(1), is commonly used to discriminate between excitation of molecular hydrogen by collisional excitation and excitation by FUV photons (e.g., \citealp{Hayashi85}; \citealp{Black87}; \citealp{Burton92}).

In the case of fluorescent excitation in a low-density PDR (n$_{\mathrm{H_{2}}}$ $<$ 5 $\times$ 10$^{4}$ cm$^{-3}$), the 2-1 S(1)/1-0 S(1) ratio is $\sim$0.56 (see, for example, Model 14 of \citealp{Black87}). In denser PDR regions or shocked regions (e.g., bipolar outflows), where collisional excitations populate the $\nu$ $=$ 0 and $\nu$ $=$ 1 levels, the ratio is $\sim$0.08 (e.g., Model S2 in \citealp{Black87}). From our data, the ratio of 2-1 S(1)/1-0 S(1) was 0.41 $\pm$ 0.01, 0.48 $\pm$ 0.02, and 0.56 $\pm$ 0.03 in regions A, B and C, respectively. The H$_{2}$ level populations in region A are somewhat more affected by collisional excitation than those in region C and the gradient in the ratios implies a gradient in the gas density.

Our measured ratios of 2-1 S(1)/1-0 S(1) are different from the case of H$_{2}$ emissions which are characterized by shock excitations (\citealp{Burton88}; \citealp{Burton89}). In the case of shocked regions, e.g., the shock driven by a supernova remnant (SNR), IC~443 \citep{Richter95}, the values of 2-1 S(1)/1-0 S(1) ratio are $\sim$0.06$-$0.2. In addition, Oh et al. (2016a, 2016b) also measured the ratio of 2-1 S(1)/1-0 S(1), $\sim$0.05$-$0.14 in LKH$\alpha$~234 and Orion KL outflow regions. From our data, the ratios of 2-1 S(1)/1-0 S(1) are larger than those in the cases of shocked regions such as IC~443, LKH$\alpha$~234 and Orion KL outflow regions.


\subsection{Ortho-to-Para Ratio}\label{opr}

Ortho H$_{2}$ (rovibrational states of odd $J$) has a total nuclear spin of 1, while para H$_{2}$ (even $J$) has a total nuclear spin of 0. The Ortho-to-Para ratio (OPR) of molecular hydrogen H$_{2}$ is the ratio of the total column density of ortho H$_{2}$ to that of para H$_{2}$, divided by the degeneracy $g_{J}$ (\citealp{Hasegawa87}; \citealp{Tanaka89}).

When H$_{2}$ is formed on the grain surface, the OPR initially has a value of 3. Radiative processes in H$_{2}$ consist of electric quadrupole transitions of $\Delta$$J$ = 0, $\pm$2. Therefore, the original OPR cannot be changed by radiative processes. However, over time, OPR can be altered by an exchange reaction of H$_{2}$ with other gas constituents such as atomic hydrogen and H$^{+}$ (\citealp{Burton92}; \citealp{Sternberg99}). The rates of this reaction are strongly dependent on the temperature. At a high temperature of T $\geq$ 200 K, the value of the OPR is expected to be 3. Otherwise, in the case of lower temperature, T $\leq$ 200 K, the OPR has lower values ($<$ 3) since more H$_{2}$ are formed in para rotational state, $J$ = 0 \citep{Sternberg99}.

Low OPRs ($<$ 1.8) has been observed in PDRs (e.g., \citealp{Hasegawa87}; \citealp{Fuente00}; \citealp{Habart11}). \citet{Sternberg99} argued that differential self-shielding of ortho and para H$_{2}$ can affect the observed OPR. Due to optical depth effects, the FUV pumping rate of ortho H$_{2}$ are reduced, and the observed OPR will be lower than the actual value \citep{Sternberg99}. In shock regions, the OPR are expected to be at or near the statistical equilibrium value of 3. But, in some cases, low OPRs ($<$ 3) are also observed (HH~54: \citealp{Neufeld98}; SNRs : \citealp{Hewitt09}). Low OPRs in shocked regions could be explained by scenarios in which these SNRs are interacting with a cold and quiescent cloud. On the other hand, the shock timescale is not long enough for the OPR of the shock excited gas to reach the equilibrium value (\citealp{Wilgenbus00}; \citealp{Hewitt09}).

Studies in the literature show that the OPR has values from 1-1.8 in H$_{2}$ fluorescent regions (e.g., \citealp{Tanaka89}; \citealp{Chrysostomou93}; \citealp{Fuente00}). Initial evidence indicated that the OPR had a value different from 3 in  NGC~2023 \citep{Hasegawa87}. These authors estimated the OPR to be 1.4$-$2.0. \citet{Tanaka89} found similar results, that OPR had a value less than 3 in PDRs. \citet{Chrysostomou93} found that OPR $=$ 1.8 $\pm$ 0.5 for the M17. \citet{Habart11} estimated an OPR of 1.8 for NGC~2023. \citet{Fuente00} found that the OPR varies from 1.5$-$3 for NGC~7023.

\begin{figure}[p]
\figurenum{6}
\centering
\includegraphics[width=75ex,height=90ex, keepaspectratio]{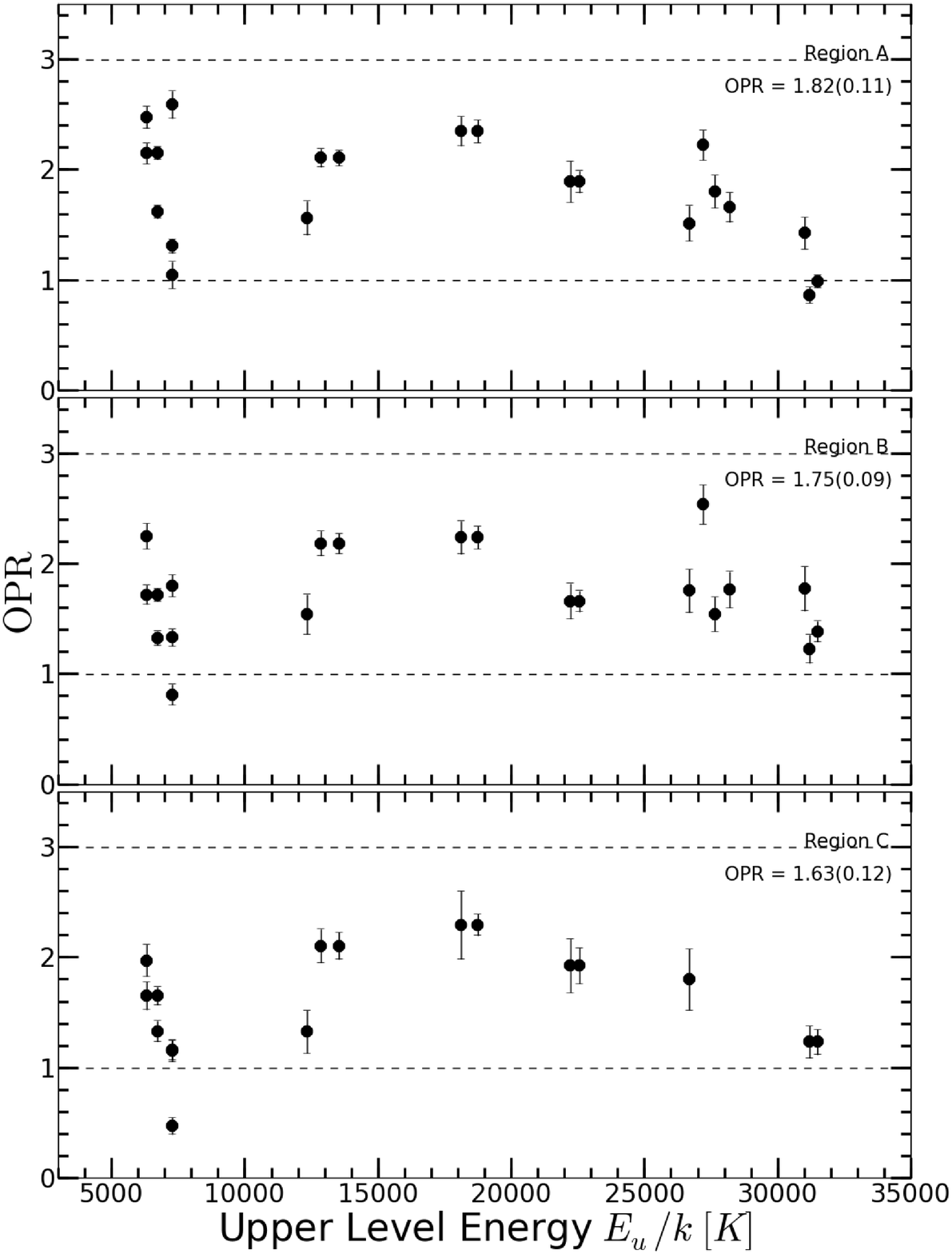}
\caption{\label{OPR} Plots of the orthor-to-para ratio (OPR) as a function of vibrational levels of regions A, B, and C. The dash-lines indicate the R$_{\mathrm{OP}}$ ratios of 1 and 3. The text in the plots indicate the average OPRs of each region.
}
\end{figure}

By using the method described in \citet{Chrysostomou93}, we derived the OPR for NGC~7023 at three regions along the IGRINS slit. Table \ref{oprtable} and Figure \ref{OPR} show the OPR as a function of the upper level energy. Figure \ref{OPR} shows the OPR values of regions A, B, and C. The average OPR was 1.82 $\pm$ 0.11, 1.75 $\pm$ 0.09, and 1.63 $\pm$ 0.12 for regions A, B, and C, respectively. Our derived OPR values are lower than that of \citet{Martini97} who presented an OPR of 2.5 $\pm$ 0.3 for the northern filament of NGC~7023. In the calculation of \citet{Martini97}, they used one pair of upper state populations of the vibrational level $\nu$ = 1 and $J$ = 3$-$4. \citet{Martini97} observed the NW part of NGC~7023 at 40$\farcs$0 west, 34$\farcs$0 north of the central star, which is offset 10$\arcsec$ to the north from ours. Based on the argument of \citep{Sternberg99}, the FUV pumping rate of para H$_{2}$ is reduced by optical depth effects. Therefore, the derived OPR of \citet{Martini97} and ours are just lower limits for the true value of the OPR. From the comparison between our derived OPR with that of \citet{Martini97}, we see that OPR varies from $\sim$1.6 to 2.5 across the PDR region.


\subsection{H$_{2}$ Column Density}

\begin{figure}[p]
\figurenum{7a}
\centering
\includegraphics[width=100ex, angle=90]{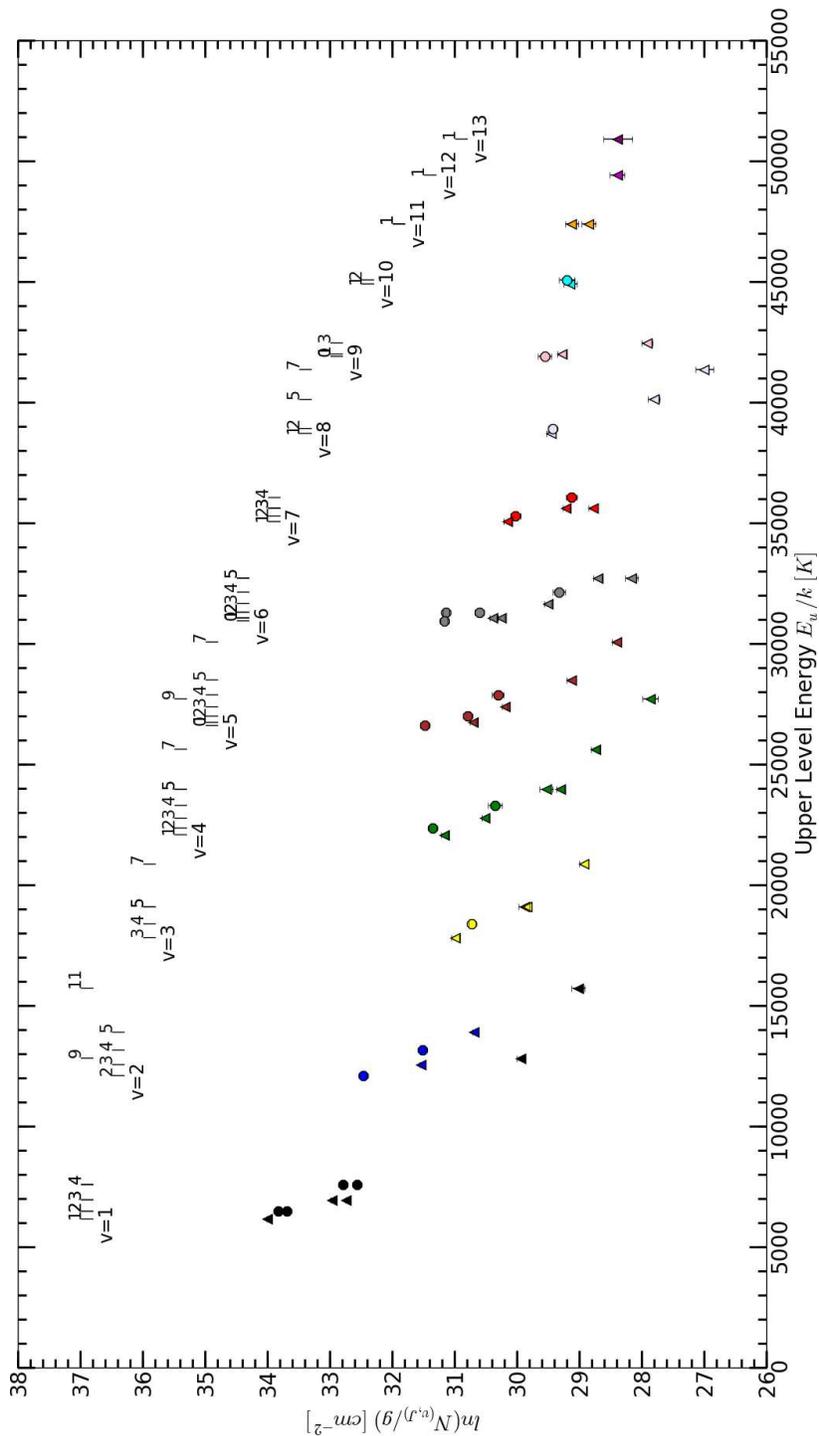}
\caption{\label{diagram1} Excitation diagram of the H$_{2}$ level column density distribution for region A. The observed column densities have been corrected for extinction, A$_{V}$ =  2.2 mag. The column densities have been divided by the level degeneracies, assuming the ortho-to-para ratio is 3. The circles and triangles present the ortho and para H$_{2}$, respectively.  
}
\end{figure}
\begin{figure}[p]
\figurenum{7b}
\centering
\includegraphics[width=100ex, angle=90]{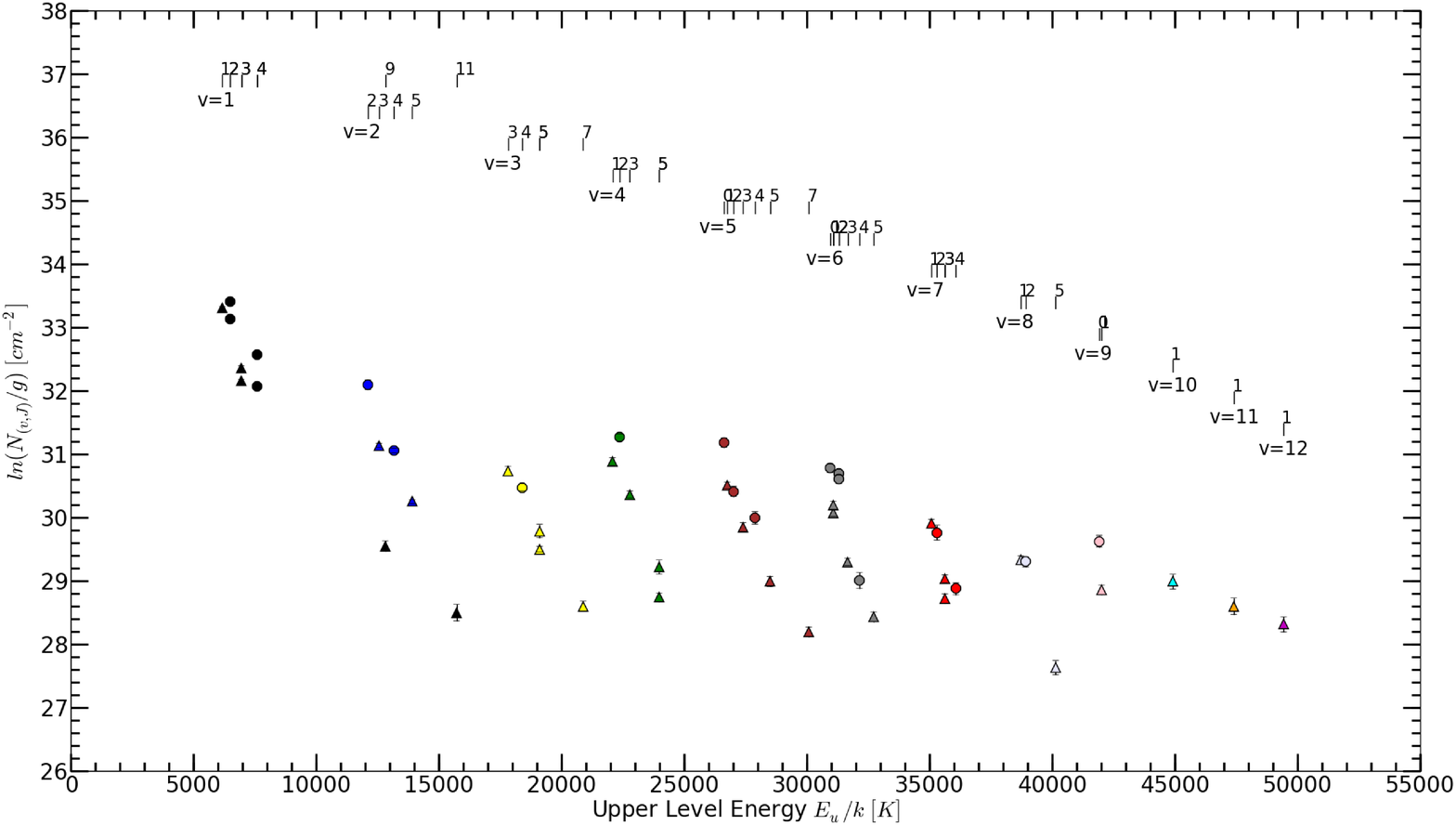}
\caption{\label{diagram2} Same as Figure 7a, except for the spectra in region B.
}
\end{figure}
\begin{figure}[p]
\figurenum{7c}
\centering
\includegraphics[width=100ex, angle=90]{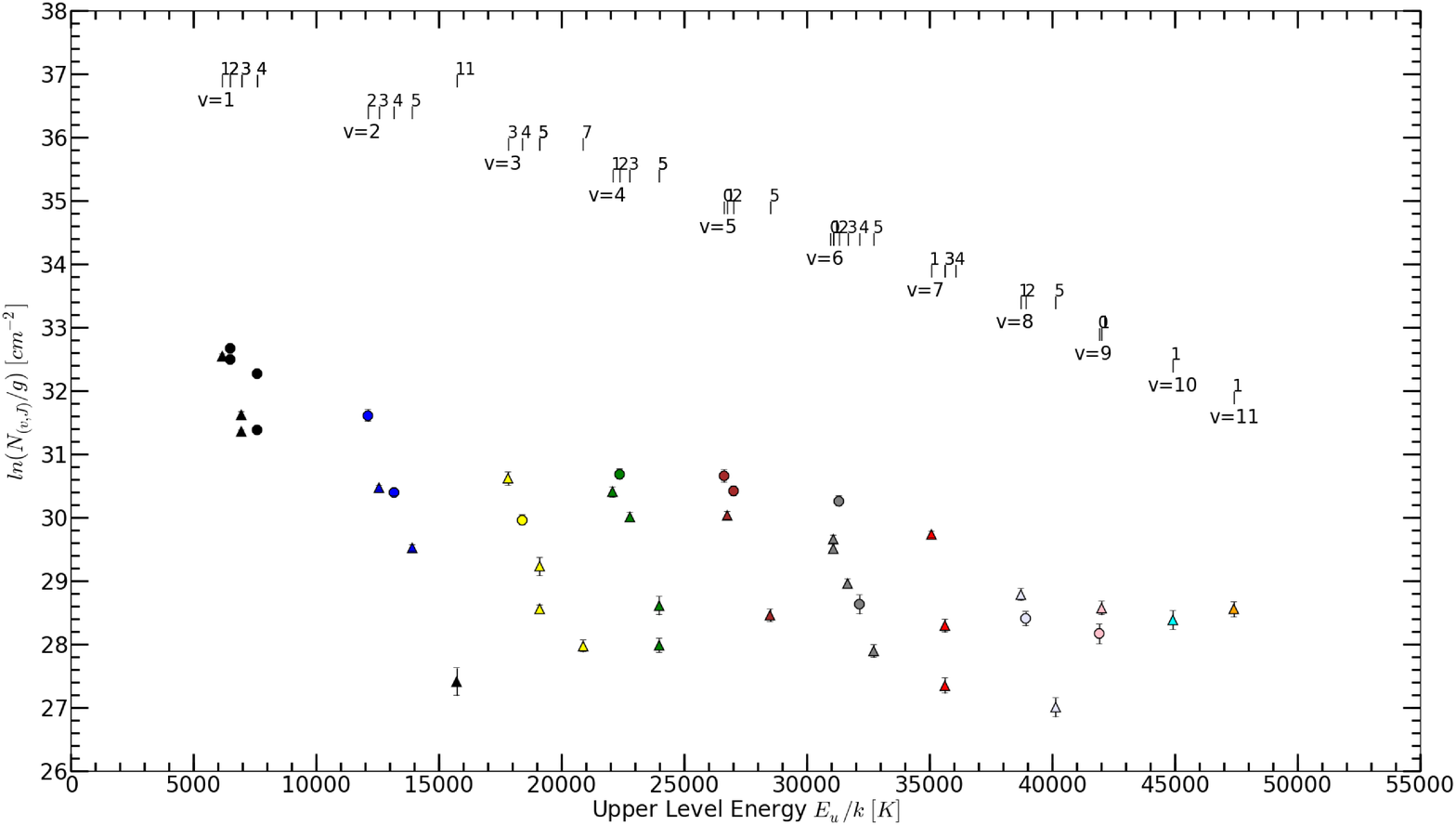}
\caption{\label{diagram3} Same as Figure 7a, except for the spectra in region C.
}
\end{figure}

Molecular H$_{2}$ emission lines are optically thin since they have small radiative transition probabilities. We present the level column densities divided by the level degeneracy $g_{J}$ against upper level energy $E_{\mathrm{u}}(\nu,J)/k$ in Table \ref{h2line}. Figures \ref{diagram1}, \ref{diagram2}, and \ref{diagram3} show the excitation diagrams of molecular H$_{2}$ emission lines in regions A, B, and C, respectively.

These excitation diagrams clearly reveal the H$_{2}$ excitation mechanism. For thermally excited H$_{2}$, the excitation diagram would show a single smooth function of $\ln$$N(\nu ,J)$ against $E_{\mathrm{u}}(\nu,J)$. In the case of fluorescent excitation, the plot shows characteristic sawtooth distribution, where vibrational temperatures are much larger than rotational temperatures but each rotational ladder is characterized by a temperature that increases with increasing $J$ (e.g., \citealp{Hasegawa87}; \citealp{Tanaka89}; \citealp{Burton92}; \citealp{Chrysostomou93}; \citealp{Ramsay93}; \citealp{Draine96}; \citealp{Rosenthal00}). Figures \ref{diagram1}$-$\ref{diagram3} show sawtooth distributions, which clearly indicate that the H$_{2}$ emission lines are mostly UV excited.

\section{Discussion}

\subsection{Comparison of the H$_{2}$ spectra to a PDR model}

To investigate the physical conditions in the observed region, we compare the observed emission lines to PDR models from the literature. The large number of H$_{2}$ emission lines in the IGRINS spectrum of NGC~7023 allows us to compare observed line strengths with the PDR models in detail. Here we compare the line ratios of all observed H$_{2}$ emission lines to the PDR model of \citet{Draine96} (hereafrer DB96). DB96 provides H$_{2}$ emission lines at densities from 10$^2$ cm$^{-3}$ to 10$^{6}$ cm$^{-3}$ and a UV field intensities from 1 {\it G$_{0}$} to 10$^{4}$ {\it G$_{0}$}. The A$_{V}$ value used in Section \ref{spectra} is the stellar extinction value, and the nebula extinction value may differ from that. \citet{Martini97,Martini99} agree that this variation of this extinction value by a factor of 2 does not affect their results. We also test how the line ratio of 2-1 S(1)/1-0 S(1) varies while changing the extinction value of A$_{V}$ by $\Delta$A$_{V}$ = $\pm$2 mag. Our results show that the line ratio varies within $\pm$0.01 with changing the extinction value, which has no meaningful effect on our calculations.

\begin{figure*}[p]
\figurenum{8}
\centering
\includegraphics[width=75ex,height=90ex, keepaspectratio]{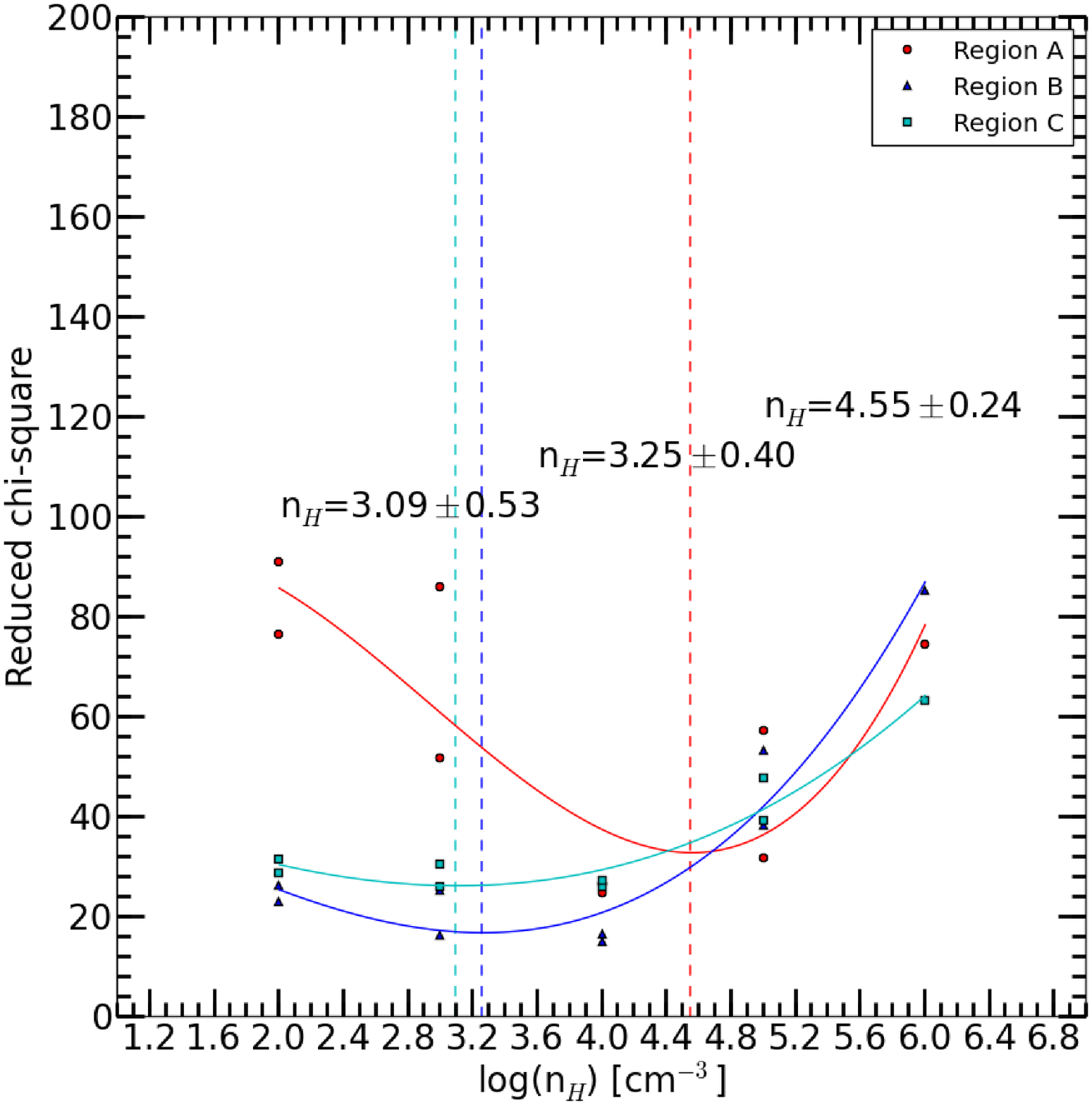}
\caption{The estimate densities of regions A (Red), B (Blue), and C (Cyan) from DB96. The solid lines display the fitting of reduced chi-square values of regions A, B, and C. The dash lines show the minimum reduced chi-square values from the fitting.
}
\label{denfit}
\end{figure*}

The best fits are determined from the minimum reduced chi-square value of the differences between the line ratios in the model and those in the data. Table \ref{pdrscom} shows the derived reduced chi-square values between the observed data and the models. Those chi-square values are fitted again with a curve to find the minimum in between the models. The best fits are at n$_{H}$ $=$ 10$^{4.55 \pm 0.24}$ cm$^{-3}$, 10$^{3.25 \pm 0.40}$ cm$^{-3}$, and 10$^{3.09 \pm 0.53}$ cm$^{-3}$ for region A, B, and C respectively. Figure \ref{denfit} shows how we derive the densities from the minimum reduced chi-square values. By using low spectral resolution NIR H$_{2}$ emission lines, \citet{Martini97,Martini99} found that the NW filament of NGC~7023 has a high density clump of n$_{H}$ $\sim$ 10$^{6}$ cm$^{-3}$ in the filament with a lower density of 10$^4$$-$10$^{5}$ cm$^{-3}$. By using radio emission lines from ISO observations, \citet{Fuente00} found that the NW filament has high density filament of n$_{H}$ $\sim$ 10$^{6}$ cm$^{-3}$ immersed in a diffuse interfilament of n$_{H}$ $\sim$ 10$^{4}$ cm$^{-3}$. \citet{Kohler14} estimated the density values for the NW filament of n$_{H}$ $=$ 10$^4$$-$10$^{6}$ cm$^{-3}$ by using CO emission from Spitzer observations.

\begin{figure}[p]
\figurenum{9a}
\centering
\includegraphics[width=100ex, angle=90]{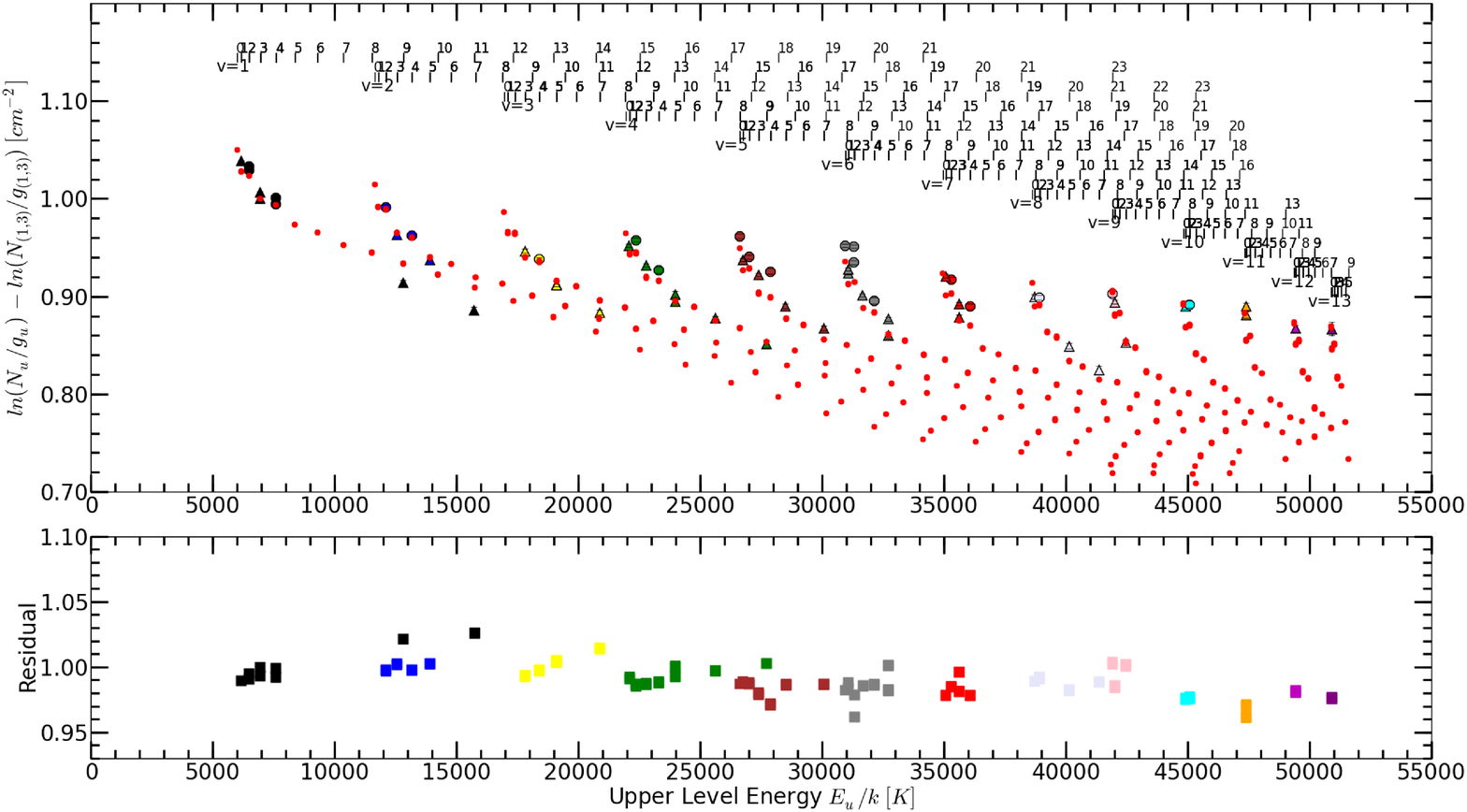}
\caption{\label{dbra} Top plot is same as Figure 7a. The red-circle markers present the model calculation of DB96, for model Lw3o, with n$_{H}$ = 10$^5$ cm$^{-3}$ and G = 10$^3$ {\it G$_{0}$}. Bottom plot displays the residual values between the model results and the observational data.
}
\end{figure}
\begin{figure}[p]
\figurenum{9b}
\centering
\includegraphics[width=100ex, angle=90]{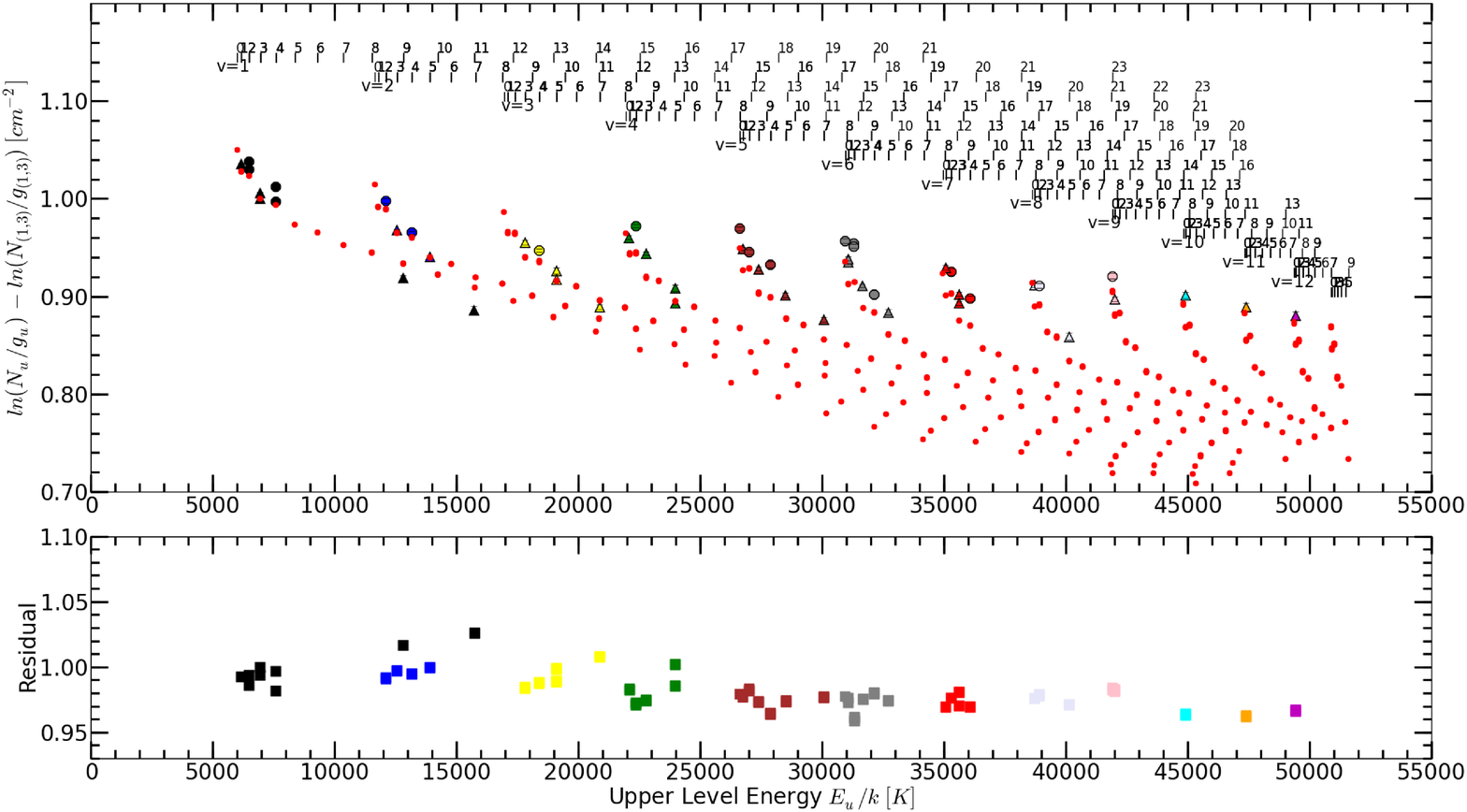}
\caption{\label{dbrb} Top plot is same as Figure 7b. The red-circle markers present the model calculation of DB96, for model Hw3o, with n$_{H}$ = 10$^4$ cm$^{-3}$ and G = 10$^3$ {\it G$_{0}$}. Bottom plot displays the residual values between the model results and the observational data.
}
\end{figure}
\begin{figure}[p]
\figurenum{9c}
\centering
\includegraphics[width=100ex, angle=90]{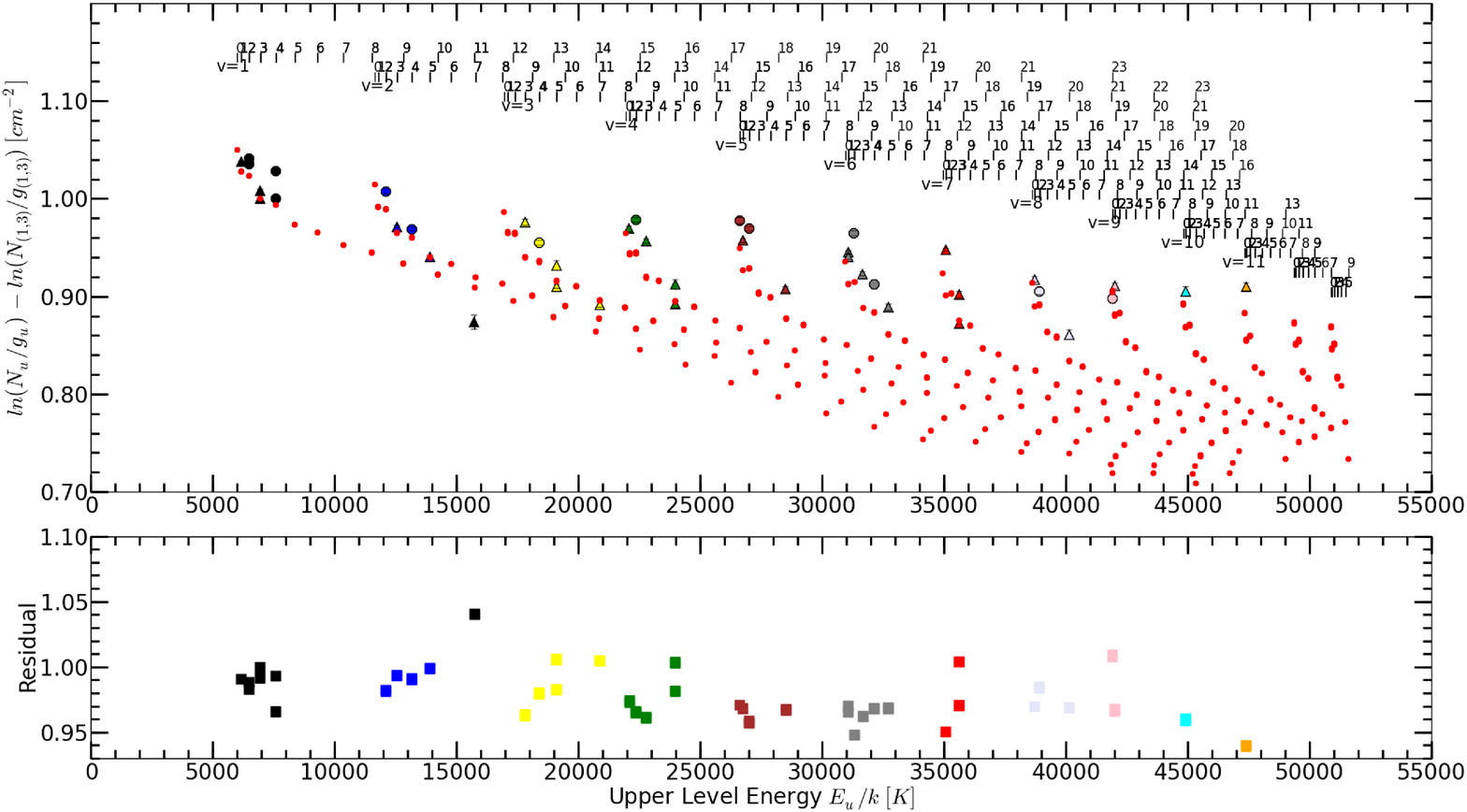}
\caption{\label{dbrc} Top plot is same as Figure 7c. The red-circle markers present the model calculation of DB96, for model Hw3o, with n$_{H}$ = 10$^4$ cm$^{-3}$ and G = 10$^3$ {\it G$_{0}$}. Bottom plot displays the residual values between the model results and the observational data.
}
\end{figure}

Figures \ref{dbra}$-$\ref{dbrc} show the normalized H$_{2}$ level column density distribution of the observational data and the calculation of model DB96 for regions A, B, and C. In region A, we display the model Lw30 of DB96, which has a density of n$_{H}$ $=$ 10$^{5}$ cm$^{-3}$. In regions B and C, we use the model Hw30, which has n$_{H}$ $=$ 10$^{4}$ cm$^{-3}$. The UV intensity of both DB96 models is G $=$ 10$^3$ {\it G$_{0}$}. Both models have densities that are close to those values, determined from the best-fit lines using the reduced chi-square method discussed above. The gas temperature assumed in the models as T$_{gas}$ $=$ 500 K is close to the temperature, we estimated using the H$_{2}$ line widths in Section \ref{h2shift}.

Figures \ref{dbra}$-$\ref{dbrc} show good agreement of the H$_{2}$ level populations between the model predictions and the observational data for all vibrational excited levels of $\nu$ $=$ 1$-$13. Although the agreement is good for all vibrational levels, there are significant differences (by factors of 2$-$3) between the model results and the observed data for the H$_{2}$ emission lines in $\nu$ $=$ 1 and rotational states $J$ $=$ 9 and 11. This discrepancy may come from the uncertainty in the rovibrational distribution function $\delta(\upsilon,J)$ of newly-formed H$_{2}$ assumed in the model. In the model of DB96, the H$_{2}$ emission lines arising from low-$J$ states are not sensitive to the distribution of newly formed H$_{2}$. However, the lines at high-$J$ levels are affected by the function of $\delta(\upsilon,J)$. The emission from high $J$ levels ($J$ $\geqslant$ 7) will increase by a factor of $\sim$2 when the formation temperature varies from 2000 K to 5 $\times$ 10$^{4}$ K. In the models of Lw30 and Hw3o, the author assumed that the newly formed H$_{2}$ has a formation temperature of T$_{f}$ $=$ 5 $\times$ 10$^{4}$ K. From the comparison between the model calculations with our observed data, we expected that the formation temperature for newly formed H$_{2}$ would be lower, around T$_{f}$ $=$ 1 $\times$ 10$^{3}$ $-$ 2 $\times$ 10$^{3}$ K.

To date, the level distribution of newly-formed H$_{2}$ is unclear. The nature of grain surface controls the properties and the formation process of newly formed H$_{2}$ molecules. Depending on the grain surface formational conditions, H$_{2}$ molecules may be formed in (low-$\nu$ and high-$J$ levels) or (high-$\nu$ and low-$J$ levels) (e.g., \citealp{Hunter78}; \citealp{Duley86}). In addition, \citet{Burton92} and \citet{Burton02} argued that they may have detected newly-formed H$_{2}$ in $\nu$ $=$ 4 and 6 levels based on the H$_{2}$ observations of the reflection nebula NGC~2023 and the northern bar of the M17. In our results, we report that the level distribution of newly-formed H$_{2}$ may be in the levels of $\nu$ $=$ 1 and $J$ $\geqslant$ 9.

\subsection{An Advancing Photodissociation Front to the Molecular Cloud}

The gradient of both the temperature and the OPR supports the idea that the PDR front is moving, as previously discussed by \citet{Fuente99} and \citet{Fuente00} based on a non-equilibrium OPR for the gas with a kinetic temperature at 400 K and a gradient of the OPR across the PDR of NGC~7023. The authors discussed that the existence of an advancing PDR front is the most likely explanation for this behavior. In that scenario, the PDR is interacting continuously via the cool gas of the molecular cloud, in which the equilibrium OPR is lower than 3. The gas that was recently incorporated into the PDR is predicted to have an OPR lower than the equilibrium value, and also has the lowest temperature. The OPR will increase and reach values close to 3 at higher temperatures.

In Section \ref{h2shift}, our results indicate a gradient in the kinetic temperature between $\sim$400 and $\sim$900 K from regions B and C to region A. In Section \ref{opr}, we found that the OPR increases from $\sim$1.6 to 2.5 across the PDR region. Moreover, \citet{Lemaire99} found a velocity gradient in the molecular species in the PDR. We confirmed an offset in the velocity of H$_{2}$ ($V_{\mathrm{LSR}}$ = 3.50 $\pm$ 0.16 km s$^{-1}$) relative to other molecular species (e.g., CO line) in the PDR region. Our results confirm the argument of \citet{Fuente99,Fuente00} that the observed region is an advancing PDR front relative to the molecular cloud.

\subsection{Size of the H$_{2}$ Clumps}

\begin{figure*}[p]
\figurenum{10}
\centering
\includegraphics[width=70ex,height=70ex, keepaspectratio]{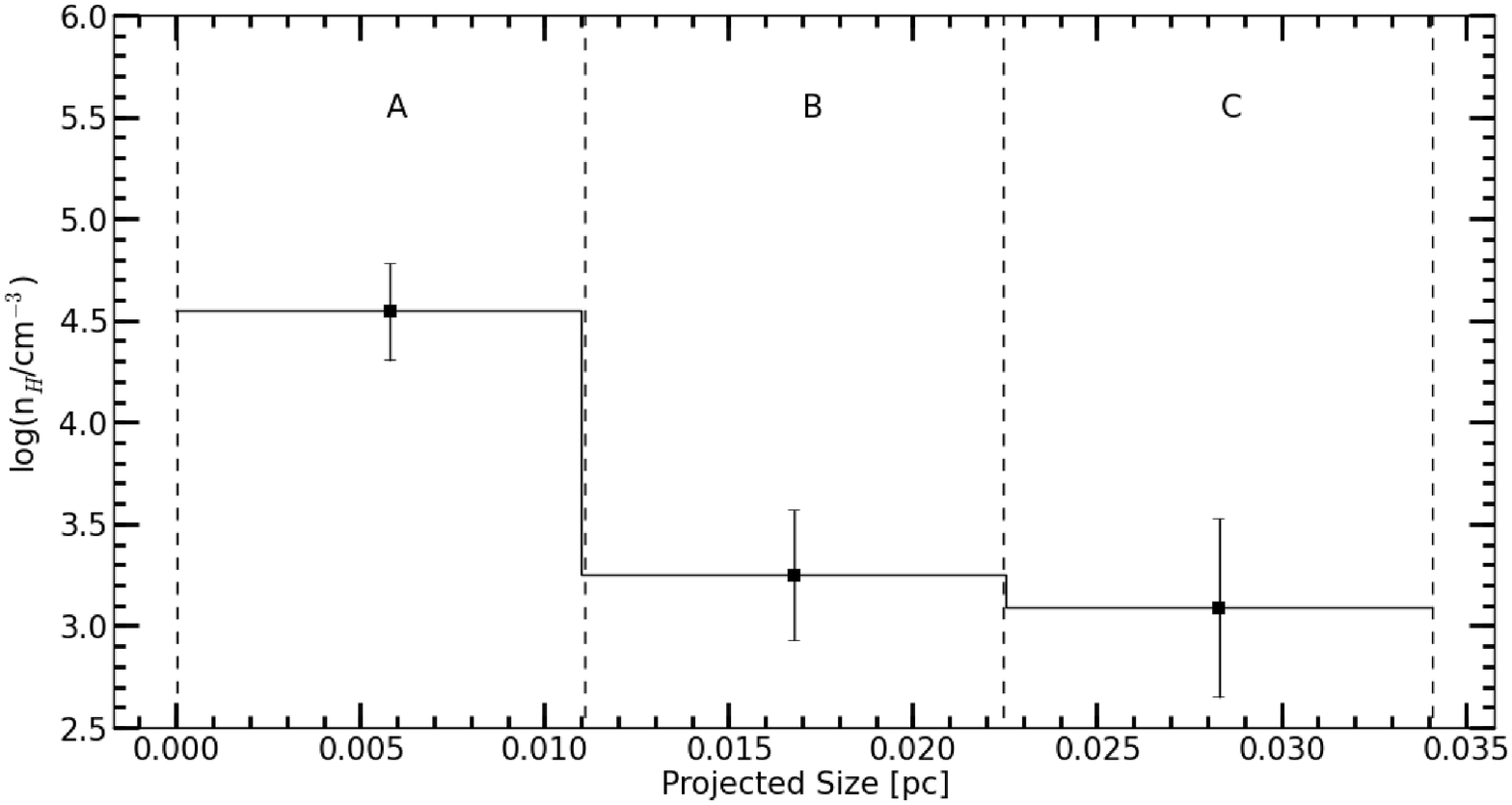}
\includegraphics[width=70ex,height=70ex, keepaspectratio]{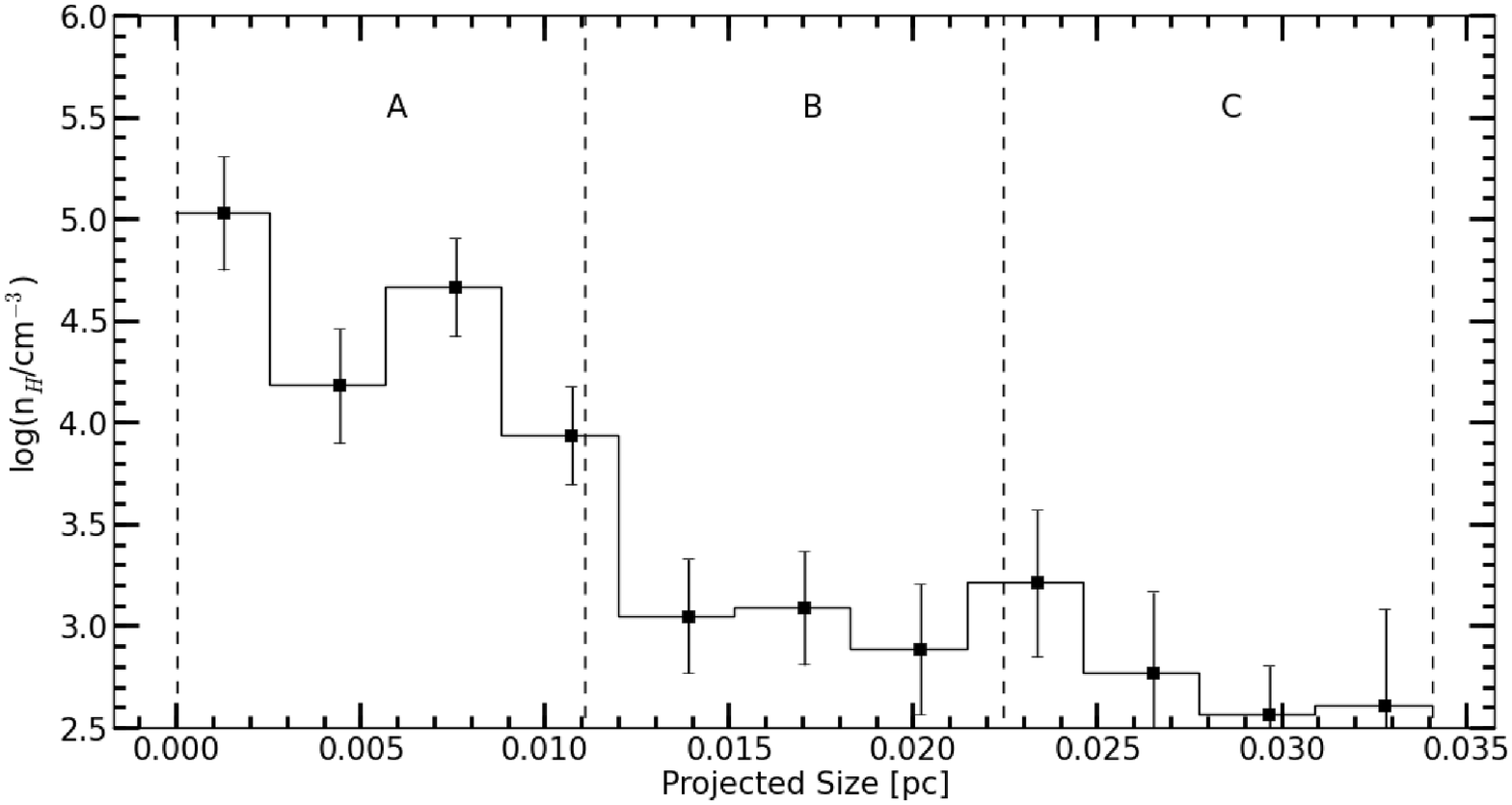}
\caption{Top-plot displays the estimated densities of regions A, B, and C (spatial resolution of $\sim$5$\arcsec$ or 0.01 pc) from the PDR model of DB96. The solid line displays the estimate densities by DB96. The bottom-plot displays the estimated densities by DB96 with a spatial resolution of $\sim$1$\arcsec$ or 0.002 pc.
}
\label{rmsd}
\end{figure*}

We found a variation in density distribution throughout regions A, B, and C within the observational area of 1$\arcsec$ $\times$ 15$\arcsec$. The gas density is higher in region A, and decreases toward regions B and C. In Figure \ref{rmsd}, the top-plot shows the density profile of regions A, B, and C (in each area of 1$\arcsec$ $\times$ 5$\arcsec$, corresponding to a spatial scale of 0.01 pc). The bottom-plot displays the estimated densities, covering an area of 1$\arcsec$ $\times$ 1$\arcsec$, corresponding to a scale of 0.002 pc. We see variations of the density within each region. In particular, the density distribution in region A varies from 10$^{5.03 \pm 0.32}$ to 10$^{4.18 \pm 0.32}$ cm$^{-3}$ within $\sim$0.01 pc. Across region A, the variation of density distribution shows two peaks of densities. One peak has a density of 10$^{5.03 \pm 0.32}$ cm$^{-3}$, and the other is 10$^{4.67 \pm 0.24}$ cm$^{-3}$. The gradients of the density distribution may indicate evidence of a high density clump with a size smaller than $\sim$5 $\times$ 10$^{-3}$ pc.

Through the variations of density distribution, we found clumpy structures in the observed area, including a high density clump of n$_{H}$ $\sim$ 10$^{5}$ cm$^{-3}$ with a size of $\sim$5 $\times$ 10$^{-3}$ pc or less, embedded in the lower density regions of n$_{H}$ $=$ 10$^{3}$$-$10$^{4}$ cm$^{-3}$. Previous works argue that the clump sizes are 10$^{-3}$ $-$ 10$^{-4}$ pc (\citealp{Garay87}; \citealp{Churchwell87}). \citet{Sellgren92} also found that NGC~7023 has a clumpy structure. The clump size that we derive is consistent with the previous report from \citet{Lemaire96}. By using a high spatial resolution 0$\farcs$8 for the H$_{2}$ 1-0 S(1) and 1-0 S(2) observations from the NW of NGC~7023, the authors found that there is evidence of small scale structures in a scale size of 3 to 4 $\times$ 10$^{-3}$ pc.

The increasing density within the small scale may also be evidence of local turbulence in the region, which has important connections to the star-formation processes. These rapid changes of density within the region of subparsec scales gives us a glimpse of where new stars are formed. A typical size scale for a region where stars are born is within ``cores'' sizes of 0.03 $-$ 0.2 pc with a mean density of 10$^{4}$$-$10$^{5}$ cm$^{-3}$ (\citealp{Kennicutt00}; \citealp{Bergin07}; \citealp{Thompson07}).

\section{Summary and Conclusion}\label{sum}

We analyzed the near-infrared H$_{2}$ emission lines from the 1$\arcsec$ $\times$ 15$\arcsec$ region in the NW filament of the reflection nebula, NGC~7023. The high spatial and spectral resolution of IGRINS provides many H$_{2}$ rovibrational emission lines, which we have used to resolve the physical structure of the PDR, and look into details of the physical excitation mechanisms of the observed regions. With these observations we have determined:

The diagnostic ratios of 2-1 S(1)/1-0 S(1) in regions A, B, and C are between 0.41$-$0.56. The ratios indicate that regions B and C are mostly UV excited, while the excitation mechanism in region A is partially collisional excitation or collisional de-excitation. The derived OPR of 1.63$-$1.82 also indicates that the observed H$_{2}$ lines are UV fluorescence. In addition, the distributions of the excitation diagrams confirm that the detected H$_{2}$ emission lines are from PDR.

The exhibits of the gradient of kinetic temperature, velocity, and OPR of H$_{2}$ emission lines in the observed areas demonstrate a presence of the dynamic PDR front relative to the molecular cloud.

We found the gradient density distribution within the observed regions. We suggest that the area has a clumpy structure, including a high density clump of $\sim$10$^{5}$ cm$^{-3}$ with a size smaller than $\sim$5 $\times$ 10$^{-3}$ pc embedded in lower density regions of 10$^{3}$$-$10$^{4}$ cm$^{-3}$ closer to the Herbig B3Ve$-$B5 star HD~200775.

\section*{Acknowledgments}

We thank the referee, Dr. Jeonghee Rho, for valuable suggestions and comments, which improved the presentation and clarify of the paper. We also thank Dr. Roueff for noticing the mis-identified emission line. This work was supported by a National Research Foundation of Korea (NRF) grant, No. 2008-0060544, funded by the Ministry of Science, ICT and Future Planning (MSIP) of Korea. Huynh Anh N. Le and Hye-In Lee were supported by the BK21 plus program through a NRF funded by the Ministry of Education of Korea. This work used the Immersion Grating Infrared Spectrograph (IGRINS), which was developed under a collaboration between the University of Texas at Austin and the Korea Astronomy and Space Science Institute (KASI) with the financial support of the US National Science Foundation under grant AST$-$1229522, of the University of Texas at Austin, and of the Korean GMT Project of KASI. The IGRINS software packages were developed by Kyung Hee University based on a contract of KASI. This paper includes data taken at The McDonald Observatory at the University of Texas at Austin.


\begin{deluxetable}{ccccccccccccc}
\rotate
\centering
\tabletypesize{\scriptsize}
\tablecolumns{12}
\tablewidth{0pt}
\tablecaption{NGC~7023 H$_{2}$ line observation \label{h2line}}
\tablehead{
\colhead{}  &  \colhead{}    &  \multicolumn{3}{c}{Region A}  & \colhead{} &  \multicolumn{3}{c}{Region B}   & \colhead{}  &  \multicolumn{3}{c}{Region C} \\
\cline{3-5} \cline{7-9} \cline{11-13} \\
\colhead{Line} & \colhead{$\lambda$\tablenotemark{a}} &  \colhead{I\tablenotemark{b}} &  \colhead{I\tablenotemark{c}}  & \colhead{$\ln(N_{u}/g)$\tablenotemark{d}} &  \colhead{}  & \colhead{I\tablenotemark{b}}  &  \colhead{I\tablenotemark{c}} & \colhead{$\ln(N_{u}/g)$\tablenotemark{d}} &  \colhead{} & \colhead{I\tablenotemark{b}} &  \colhead{I\tablenotemark{c}}  &  \colhead{$\ln(N_{u}/g)$\tablenotemark{d}}}

\startdata
4-2 Q(9)	&	1.498884	&	0.863(0.107)	&	0.041(0.005)	&	27.867(0.124)	&	&	\nodata	&	\nodata	&	\nodata	&	&	\nodata	&	\nodata	&	\nodata	\\
4-2 O(3)	&	1.509865	&	5.782(0.311)	&	0.275(0.015)	&	31.156(0.054)	&	&	4.432(0.269)	&	0.372(0.024)	&	30.890(0.061)	&	&	2.731(0.224)	&	0.511(0.044)	&	30.406(0.082)	\\
5-3 Q(4)	&	1.515792	&	2.356(0.229)	&	0.112(0.011)	&	30.306(0.097)	&	&	1.737(0.168)	&	0.146(0.014)	&	30.001(0.096)	&	&	\nodata	&	\nodata	&	\nodata	\\
3-1 O(5)	&	1.522026	&	2.920(0.188)	&	0.139(0.009)	&	30.990(0.064)	&	&	2.271(0.172)	&	0.190(0.015)	&	30.739(0.076)	&	&	2.022(0.212)	&	0.378(0.041)	&	30.623(0.105)	\\
5-3 Q(5)	&	1.528641	&	2.562(0.199)	&	0.122(0.010)	&	29.125(0.078)	&	&	2.251(0.186)	&	0.189(0.016)	&	28.996(0.083)	&	&	1.320(0.131)	&	0.247(0.025)	&	28.462(0.099)	\\
6-4 S(0)	&	1.536908	&	1.942(0.150)	&	0.092(0.007)	&	30.600(0.077)	&	&	2.134(0.156)	&	0.179(0.013)	&	30.694(0.073)	&	&	\nodata	&	\nodata	&	\nodata	\\
10-7 O(3)	&	1.548851	&	1.039(0.111)	&	0.049(0.005)	&	29.146(0.106)	&	&	0.893(0.102)	&	0.075(0.009)	&	28.995(0.114)	&	&	0.488(0.072)	&	0.091(0.014)	&	28.390(0.148)	\\
5-3 O(2)	&	1.560730	&	2.524(0.169)	&	0.120(0.008)	&	31.481(0.067)	&	&	1.876(0.138)	&	0.157(0.012)	&	31.184(0.073)	&	&	1.113(0.114)	&	0.208(0.022)	&	30.662(0.103)	\\
5-3 Q(7)	&	1.562627	&	1.578(0.120)	&	0.075(0.006)	&	28.402(0.076)	&	&	1.285(0.098)	&	0.108(0.008)	&	28.196(0.076)	&	&	\nodata	&	\nodata	&	\nodata	\\
4-2 O(4)	&	1.563516	&	2.546(0.160)	&	0.121(0.008)	&	31.355(0.063)	&	&	2.340(0.139)	&	0.196(0.012)	&	31.271(0.059)	&	&	1.311(0.110)	&	0.245(0.022)	&	30.692(0.084)	\\
7-5 S(2)	&	1.588290	&	1.240(0.113)	&	0.059(0.005)	&	29.124(0.091)	&	&	0.973(0.097)	&	0.082(0.008)	&	28.881(0.100)	&	&	0.860(\nodata)	&	0.161(\nodata)	&	28.758(\nodata)	\\
6-4 Q(1)	&	1.601535	&	3.992(0.193)	&	0.190(0.010)	&	30.248(0.048)	&	&	3.369(0.173)	&	0.283(0.015)	&	30.078(0.051)	&	&	2.222(0.134)	&	0.416(0.028)	&	29.662(0.060)	\\
6-4 Q(2)	&	1.607386	&	3.813(0.194)	&	0.182(0.010)	&	31.136(0.051)	&	&	2.255(0.157)	&	0.189(0.014)	&	30.611(0.070)	&	&	1.600(0.138)	&	0.299(0.027)	&	30.268(0.086)	\\
5-3 O(3)	&	1.613536	&	4.932(0.237)	&	0.235(0.012)	&	30.697(0.048)	&	&	4.112(0.214)	&	0.345(0.019)	&	30.515(0.052)	&	&	2.549(0.165)	&	0.477(0.034)	&	30.037(0.065)	\\
13-9 Q(1)	&	1.614812	&	0.247(0.057)	&	0.012(0.003)	&	28.383(0.233)	&	&	\nodata	&	\nodata	&	\nodata	&	&	\nodata	&	\nodata	&	\nodata	\\
6-4 Q(3)	&	1.616211	&	2.879(0.158)	&	0.137(0.008)	&	29.505(0.055)	&	&	2.353(0.136)	&	0.197(0.012)	&	29.303(0.058)	&	&	1.673(0.116)	&	0.313(0.023)	&	28.963(0.069)	\\
7-5 S(1)	&	1.620530	&	2.723(0.145)	&	0.130(0.007)	&	29.208(0.053)	&	&	2.283(0.143)	&	0.191(0.013)	&	29.032(0.063)	&	&	1.098(0.116)	&	0.205(0.022)	&	28.300(0.106)	\\
4-2 O(5)	&	1.622292	&	3.133(0.162)	&	0.149(0.008)	&	30.507(0.052)	&	&	2.716(0.154)	&	0.228(0.014)	&	30.364(0.057)	&	&	1.914(0.143)	&	0.358(0.029)	&	30.014(0.075)	\\
6-4 Q(4)	&	1.628084	&	0.982(0.098)	&	0.047(0.005)	&	29.325(0.100)	&	&	0.718(0.093)	&	0.060(0.008)	&	29.012(0.129)	&	&	0.492(0.074)	&	0.092(0.014)	&	28.634(0.151)	\\
6-4 Q(5)	&	1.643080	&	1.858(0.129)	&	0.088(0.006)	&	28.703(0.070)	&	&	1.425(0.115)	&	0.119(0.010)	&	28.438(0.081)	&	&	0.832(0.084)	&	0.156(0.016)	&	27.900(0.101)	\\
3-1 O(7)	&	1.645324	&	0.668(0.068)	&	0.032(0.003)	&	29.871(0.102)	&	&	0.615(0.065)	&	0.052(0.006)	&	29.789(0.106)	&	&	0.353(0.051)	&	0.066(0.010)	&	29.233(0.145)	\\
11-8 Q(1)	&	1.657105	&	0.730(0.083)	&	0.035(0.004)	&	28.851(0.114)	&	&	0.570(0.075)	&	0.048(0.006)	&	28.604(0.131)	&	&	0.545(0.066)	&	0.102(0.013)	&	28.558(0.122)	\\
7-5 S(0)	&	1.658482	&	1.054(0.087)	&	0.050(0.004)	&	30.020(0.083)	&	&	0.818(0.095)	&	0.069(0.008)	&	29.767(0.116)	&	&	\nodata	&	\nodata	&	\nodata	\\
8-6 S(5)	&	1.664578	&	0.518(0.075)	&	0.025(0.004)	&	26.994(0.144)	&	&	\nodata	&	\nodata	&	\nodata	&	&	\nodata	&	\nodata	&	\nodata	\\
5-3 O(4)	&	1.671821	&	1.999(0.138)	&	0.095(0.007)	&	30.793(0.069)	&	&	1.372(0.114)	&	0.115(0.010)	&	30.417(0.083)	&	&	1.382(0.107)	&	0.259(0.021)	&	30.424(0.077)	\\
6-4 O(2)	&	1.675019	&	2.127(0.128)	&	0.101(0.006)	&	31.167(0.060)	&	&	1.451(0.102)	&	0.122(0.009)	&	30.784(0.071)	&	&	1.125(\nodata)	&	0.211(\nodata)	&	30.530(\nodata)	\\
4-2 O(6)	&	1.686494	&	0.791(0.085)	&	0.038(0.004)	&	30.350(0.108)	&	&	\nodata	&	\nodata	&	\nodata	&	&	\nodata	&	\nodata	&	\nodata	\\
1-0 S(9)	&	1.687721	&	1.017(0.108)	&	0.048(0.005)	&	29.014(0.106)	&	&	0.610(0.081)	&	0.051(0.007)	&	28.504(0.133)	&	&	0.205(0.045)	&	0.038(0.008)	&	27.415(0.218)	\\
8-6 S(3)	&	1.701797	&	1.005(0.099)	&	0.048(0.005)	&	27.806(0.098)	&	&	0.846(0.093)	&	0.071(0.008)	&	27.635(0.110)	&	&	0.453(0.068)	&	0.085(0.013)	&	27.010(0.150)	\\
7-5 Q(1)	&	1.728779	&	3.613(0.211)	&	0.172(0.010)	&	30.150(0.058)	&	&	2.836(0.207)	&	0.238(0.018)	&	29.907(0.073)	&	&	2.394(0.129)	&	0.448(0.027)	&	29.738(0.054)	\\
6-4 O(3)	&	1.732637	&	4.227(0.233)	&	0.201(0.011)	&	30.382(0.055)	&	&	3.519(0.212)	&	0.295(0.019)	&	30.199(0.060)	&	&	1.776(0.129)	&	0.332(0.026)	&	29.515(0.073)	\\
5-3 O(5)	&	1.735888	&	3.222(0.159)	&	0.153(0.008)	&	30.194(0.049)	&	&	2.298(0.154)	&	0.193(0.013)	&	29.856(0.067)	&	&	\nodata	&	\nodata	&	\nodata	\\
1-0 S(7)	&	1.748035	&	3.624(0.218)	&	0.173(0.011)	&	29.937(0.060)	&	&	2.475(0.204)	&	0.208(0.018)	&	29.555(0.082)	&	&	\nodata	&	\nodata	&	\nodata	\\
4-2 O(7)	&	1.756296	&	0.873(0.091)	&	0.042(0.004)	&	29.527(0.105)	&	&	0.645(0.070)	&	0.054(0.006)	&	29.225(0.108)	&	&	0.351(0.052)	&	0.066(0.010)	&	28.617(0.148)	\\
11-8 O(3)	&	1.760929	&	0.940(0.100)	&	0.045(0.005)	&	29.124(0.106)	&	&	\nodata	&	\nodata	&	\nodata	&	&	\nodata	&	\nodata	&	\nodata	\\
7-5 O(5)	&	2.022040	&	1.022(0.078)	&	0.049(0.004)	&	28.774(0.076)	&	&	0.972(0.073)	&	0.082(0.006)	&	28.724(0.075)	&	&	0.247(0.031)	&	0.046(0.006)	&	27.356(0.124)	\\
6-4 O(7)	&	2.029694	&	0.415(0.043)	&	0.020(0.002)	&	28.161(0.104)	&	&	\nodata	&	\nodata	&	\nodata	&	&	\nodata	&	\nodata	&	\nodata	\\
1-0 S(2)	&	2.033756	&	9.052(0.224)	&	0.431(0.012)	&	32.560(0.025)	&	&	5.579(0.174)	&	0.468(0.017)	&	32.076(0.031)	&	&	2.795(0.118)	&	0.523(0.026)	&	31.385(0.042)	\\
8-6 O(3)	&	2.041816	&	1.613(0.113)	&	0.077(0.005)	&	29.450(0.070)	&	&	1.436(0.103)	&	0.120(0.009)	&	29.333(0.072)	&	&	0.835(0.077)	&	0.156(0.015)	&	28.790(0.093)	\\
3-2 S(5)	&	2.065557	&	1.335(0.096)	&	0.064(0.005)	&	28.924(0.072)	&	&	0.971(0.078)	&	0.081(0.007)	&	28.606(0.081)	&	&	0.518(0.049)	&	0.097(0.010)	&	27.978(0.095)	\\
12-9 O(3)	&	2.069969	&	0.311(0.037)	&	0.015(0.002)	&	28.392(0.118)	&	&	0.289(0.035)	&	0.024(0.003)	&	28.321(0.122)	&	&	\nodata	&	\nodata	&	\nodata	\\
2-1 S(3)	&	2.073510	&	7.288(0.179)	&	0.347(0.010)	&	30.693(0.025)	&	&	4.718(0.141)	&	0.396(0.014)	&	30.258(0.030)	&	&	2.258(0.109)	&	0.423(0.023)	&	29.521(0.048)	\\
9-7 Q(3)	&	2.100659	&	0.387(0.034)	&	0.018(0.002)	&	27.915(0.088)	&	&	\nodata	&	\nodata	&	\nodata	&	&	\nodata	&	\nodata	&	\nodata	\\
8-6 O(4)	&	2.121570	&	0.610(0.044)	&	0.029(0.002)	&	29.429(0.072)	&	&	0.540(0.045)	&	0.045(0.004)	&	29.307(0.083)	&	&	0.220(0.025)	&	0.041(0.005)	&	28.409(0.114)	\\
1-0 S(1)	&	2.121831	&	21.002(0.291)	&	1.000	&	32.734(0.014)	&	&	11.924(0.219)	&	1.000	&	32.168(0.018)	&	&	5.343(0.146)	&	1.000	&	31.366(0.027)	\\
2-1 S(2)	&	2.154225	&	4.197(0.137)	&	0.200(0.007)	&	31.509(0.033)	&	&	2.683(0.114)	&	0.225(0.010)	&	31.061(0.043)	&	&	1.388(0.081)	&	0.260(0.017)	&	30.402(0.058)	\\
9-7 O(2)	&	2.172704	&	0.317(0.035)	&	0.015(0.002)	&	29.556(0.109)	&	&	0.341(0.032)	&	0.029(0.003)	&	29.627(0.095)	&	&	0.079(0.012)	&	0.015(0.002)	&	28.169(0.156)	\\
4-3 S(5)	&	2.200951	&	0.748(0.048)	&	0.036(0.002)	&	28.743(0.064)	&	&	\nodata	&	\nodata	&	\nodata	&	&	\nodata	&	\nodata	&	\nodata	\\
3-2 S(3)	&	2.201397	&	2.869(0.107)	&	0.137(0.005)	&	29.841(0.037)	&	&	2.043(0.091)	&	0.171(0.008)	&	29.501(0.045)	&	&	0.797(0.056)	&	0.149(0.011)	&	28.560(0.070)	\\
1-0 S(0)	&	2.223299	&	8.990(0.199)	&	0.428(0.011)	&	33.685(0.022)	&	&	5.215(0.152)	&	0.437(0.015)	&	33.140(0.029)	&	&	2.736(0.108)	&	0.512(0.025)	&	32.495(0.040)	\\
2-1 S(1)	&	2.247721	&	8.546(0.213)	&	0.407(0.012)	&	31.532(0.025)	&	&	5.767(0.174)	&	0.484(0.017)	&	31.139(0.030)	&	&	2.974(0.121)	&	0.557(0.027)	&	30.477(0.041)	\\
9-7 O(3)	&	2.253719	&	1.102(0.073)	&	0.052(0.004)	&	29.279(0.067)	&	&	0.725(0.055)	&	0.061(0.005)	&	28.861(0.076)	&	&	0.547(0.059)	&	0.102(0.011)	&	28.578(0.107)	\\
3-2 S(2)	&	2.287026	&	1.813(0.096)	&	0.086(0.005)	&	30.719(0.053)	&	&	1.424(0.080)	&	0.119(0.007)	&	30.477(0.056)	&	&	0.854(0.067)	&	0.160(0.013)	&	29.967(0.078)	\\
10-8 Q(2)	&	2.337306	&	0.233(0.029)	&	0.011(0.001)	&	29.200(0.124)	&	&	\nodata	&	\nodata	&	\nodata	&	&	\nodata	&	\nodata	&	\nodata	\\
4-3 S(3)	&	2.344479	&	1.273(0.076)	&	0.061(0.004)	&	29.298(0.060)	&	&	0.733(0.053)	&	0.061(0.005)	&	28.745(0.072)	&	&	0.345(0.039)	&	0.065(0.007)	&	27.992(0.112)	\\
2-1 S(0)	&	2.355629	&	3.663(0.224)	&	0.174(0.011)	&	32.468(0.061)	&	&	2.545(0.187)	&	0.213(0.016)	&	32.104(0.073)	&	&	1.564(0.148)	&	0.293(0.029)	&	31.617(0.095)	\\
1-0 Q(1)	&	2.406594	&	34.941(0.471)	&	1.664(0.032)	&	34.005(0.013)	&	&	17.513(0.331)	&	1.469(0.039)	&	33.314(0.019)	&	&	8.191(0.235)	&	1.533(0.061)	&	32.554(0.029)	\\
1-0 Q(2)	&	2.413436	&	11.449(0.317)	&	0.545(0.017)	&	33.827(0.028)	&	&	7.547(0.247)	&	0.633(0.024)	&	33.410(0.033)	&	&	3.608(0.175)	&	0.675(0.038)	&	32.672(0.048)	\\
1-0 Q(3)	&	2.423731	&	18.537(0.476)	&	0.883(0.026)	&	32.964(0.026)	&	&	10.166(0.352)	&	0.853(0.033)	&	32.363(0.035)	&	&	4.886(0.246)	&	0.914(0.052)	&	31.630(0.050)	\\
1-0 Q(4)	&	2.437491	&	6.285(0.212)	&	0.299(0.011)	&	32.783(0.034)	&	&	5.074(0.182)	&	0.426(0.017)	&	32.569(0.036)	&	&	3.774(0.181)	&	0.706(0.039)	&	32.273(0.048)	
\enddata
\tablenotetext{a}{The wavelengths are in units of $\mu$m.}
\tablenotetext{b}{The intensity has been corrected for extinction in units of 10$^{-19}$ W m$^{-2}$ arcsec$^{-2}$. Uncertainties are written in parentheses.}
\tablenotetext{c}{The normalized intensity with respect to the intensity of 1-0 S(1). Uncertainties are given in parentheses.}
\tablenotetext{d}{Extinction corrected upper level column density in units of $\ln$(cm$^{-2})$.}
\end{deluxetable}

\clearpage

\clearpage

\begin{deluxetable}{cccccccccccccccccc}
\rotate
\tabletypesize{\scriptsize}
\tablecolumns{12}
\tablewidth{0pc}
\tablecaption{Dynamical Information from H$_{2}$ Lines \label{dynamic}}
\tablehead{
\colhead{}    &  \multicolumn{5}{c}{Region A}  & \colhead{} &  \multicolumn{5}{c}{Region B}   & \colhead{}  &  \multicolumn{5}{c}{Region C} \\
\cline{2-6} \cline{8-12} \cline{14-18} \\
\colhead{Line} &   \colhead{$V_{shift}$\tablenotemark{a}} &  \colhead{$V_{LSR}$\tablenotemark{b}}  & \colhead{$\Delta V_{obs}$\tablenotemark{c}} &  \colhead{$\Delta V_{inst}$\tablenotemark{d}} & \colhead{$\Delta V_{fwhm}$\tablenotemark{e}} &   & \colhead{$V_{shift}$\tablenotemark{a}} &  \colhead{$V_{LSR}$\tablenotemark{b}}  & \colhead{$\Delta V_{obs}$\tablenotemark{c}} &  \colhead{$\Delta V_{inst}$\tablenotemark{d}} & \colhead{$\Delta V_{fwhm}$\tablenotemark{e}} &  & \colhead{$V_{shift}$\tablenotemark{a}} &  \colhead{$V_{LSR}$\tablenotemark{b}}  & \colhead{$\Delta V_{obs}$\tablenotemark{c}} &  \colhead{$\Delta V_{inst}$\tablenotemark{d}} & \colhead{$\Delta V_{fwhm}$\tablenotemark{e}}
\\
&   \colhead{km $s^{-1}$} &  \colhead{km $s^{-1}$}  &  \colhead{km $s^{-1}$}  &  \colhead{km $s^{-1}$}  &  \colhead{km $s^{-1}$}  &  & \colhead{km $s^{-1}$}  &  \colhead{km $s^{-1}$}  &  \colhead{km $s^{-1}$}  &  \colhead{km $s^{-1}$}  &  \colhead{km $s^{-1}$}  &  & \colhead{km $s^{-1}$}  &  \colhead{km $s^{-1}$}  &  \colhead{km $s^{-1}$}  &  \colhead{km $s^{-1}$}  &  \colhead{km $s^{-1}$}}

\startdata
4-2 O(3)	&	-18.63	&	3.90	&	. . .	&	. . .	&	. . .	&	&	-18.69	&	3.83	&	. . .	&	. . .	&	. . .	&	&	-18.88	&	3.64	&	7.20	&	7.02	&	1.62	 \\
5-3 Q(4)	&	-18.47	&	4.05	&	. . .	&	. . .	&	. . .	&	&	-19.02	&	3.50	&	8.10	&	7.04	&	4.01	&	&	-18.44	&	4.08	&	8.59	&	7.04	&	. . .    \\
3-1 O(5)	&	-17.57	&	4.96	&	. . .	&	. . .	&	. . .	&	&	. . .	&	. . .	&	. . .	&	. . .	&	. . .	&	&	-19.04	&	3.48	&	. . .	&	. . .	&	. . .	 \\
10-7 O(3)	&	-19.24	&	3.28	&	. . .	&	. . .	&	. . .	&	&	-19.65	&	2.87	&	. . .	&	. . .	&	. . .	&	&	-20.28	&	2.25	&	. . .	&	. . .	&	. . .	 \\
4-2 O(4)	&	-19.17	&	3.35	&	. . .	&	. . .	&	. . .	&	&	-19.13	&	3.39	&	. . .	&	. . .	&	. . .	&	&	. . .	&	. . .	&	. . .	&	. . .	&	. . .	 \\
6-4 Q(1)	&	-18.96	&	3.56	&	. . .	&	. . .	&	. . .	&	&	-19.00	&	3.52	&	. . .	&	. . .	&	. . .	&	&	-18.61	&	3.92	&	7.89	&	7.37	&	2.82	 \\
6-4 Q(5)	&	-18.66	&	3.86	&	8.80	&	7.54	&	4.53	&	&	-18.60	&	3.92	&	8.90	&	7.54	&	4.72	&	&	-18.80	&	3.72	&	8.55	&	7.54	&	4.04	 \\
11-8 Q(1)	&	-18.74	&	3.78	&	. . .	&	. . .	&	. . .	&	&	-18.86	&	3.66	&	8.03	&	7.60	&	2.59	&	&	. . .	&	. . .	&	. . .	&	. . .	&	. . .	 \\
5-3 O(4)	&	-20.28	&	2.24	&	. . .	&	. . .	&	. . .	&	&	-18.80	&	3.73	&	. . .	&	. . .	&	. . .	&	&	-19.87	&	2.65	&	7.98	&	7.66	&	2.24	 \\
11-8 Q(3)	&	-18.05	&	4.47	&	. . .	&	. . .	&	. . .	&	&	-18.93	&	3.59	&	. . .	&	. . .	&	. . .	&	&	. . .	&	. . .	&	. . .	&	. . .	&	. . .	 \\
8-6 S(3)	&	-17.51	&	5.01	&	8.91	&	7.80	&	4.31	&	&	. . .	&	. . .	&	. . .	&	. . .	&	. . .	&	&	. . .	&	. . .	&	. . .	&	. . .	&	. . .	 \\
8-6 S(2)	&	-17.90	&	4.63	&	9.26	&	7.92	&	4.80	&	&	. . .	&	. . .	&	. . .	&	. . .	&	. . .	&	&	. . .	&	. . .	&	. . .	&	. . .	&	. . .	 \\
5-3 O(5)	&	-18.84	&	3.68	&	8.09	&	7.95	&	1.47	&	&	-18.80	&	3.73	&	. . .	&	. . .	&	. . .	&	&	. . .	&	. . .	&	. . .	&	. . .	&	. . .	 \\
1-0 S(2)	&	-18.19	&	4.33	&	9.32	&	6.97	&	6.18	&	&	. . .	&	. . .	&	. . .	&	. . .	&	. . .	&	&	-17.96	&	4.57	&	. . .	&	. . .	&	. . .	 \\
9-7 Q(3)	&	-18.87	&	3.66	&	. . .	&	. . .	&	. . .	&	&	-19.86	&	2.67	&	7.14	&	6.89	&	. . . 	&	&	-20.46	&	2.06	&	. . .	&	. . .	&	. . .	 \\
1-0 S(1)	&	-19.02	&	3.51	&	8.53	&	6.87	&	5.06	&	&	-19.15	&	3.37	&	7.72	&	6.87	&	3.52	&	&	-19.02	&	3.50	&	7.70	&	6.87	&	3.49	 \\
9-7 O(2)	&	-17.82	&	4.70	&	8.36	&	6.81	&	4.85	&	&	-18.55	&	3.97	&	7.48	&	6.81	&	3.09	&	&	. . .	&	. . .	&	. . .	&	. . .	&	. . .	 \\
3-2 S(3)	&	-19.84	&	2.68	&	8.37	&	6.77	&	4.93	&	&	. . .	&	. . .	&	. . .	&	. . .	&	. . .	&	&	-19.99	&	2.53	&	8.03	&	6.77	&	4.31	 \\
2-1 S(1)	&	-20.15	&	2.38	&	7.62	&	6.72	&	3.59	&	&	-19.99	&	2.54	&	7.30	&	6.72	&	2.84	&	&	-19.63	&	2.89	&	7.59	&	6.72	&	3.53	 \\
9-7 O(3)	&	-19.02	&	3.50	&	. . .	&	. . .	&	. . .	&	&	-19.50	&	3.02	&	7.11	&	6.71	&	2.35	&	&	-19.87	&	2.66	&	. . .	&	. . .	&	. . .	 \\
3-2 S(2)	&	. . .	&	. . .	&	. . .	&	. . .	&	. . .	&	&	. . .	&	. . .	&	. . .	&	. . .	&	. . .	&	&	-18.76	&	3.77	&	7.31	&	6.68	&	2.97	 \\
1-0 Q(1)	&	-19.38	&	3.15	&	8.48	&	6.55	&	5.39	&	&	. . .	&	. . .	&	. . .	&	. . .	&	. . .	&	&	. . .	&	. . .	&	. . .	&	. . .	&	. . .	 \\
1-0 Q(2)	&	-19.67	&	2.85	&	8.98	&	6.54	&	6.16	&	&	. . .	&	. . .	&	. . .	&	. . .	&	. . .	&	&	. . .	&	. . .	&	. . .	&	. . .	&	. . .	 \\
1-0 Q(3)	&	-20.13	&	2.40	&	7.82	&	6.53	&	4.31	&	&	-18.42	&	4.10	&	7.27	&	6.53	&	3.19	&	&	-17.34	&	5.18	&	7.56	&	6.53	&	3.81	 \\
1-0 Q(4)	&	-19.19	&	3.33	&	. . .	&	. . .	&	. . .	&	&	. . .	&	. . .	&	. . .	&	. . .	&	. . .	&	&	. . .	&	. . .	&	. . .	&	. . .	&	. . .

\enddata
\tablenotetext{a}{The velocity shift measured from the best-fit Gaussian.}
\tablenotetext{b}{The velocity shift corrected by the earth and heliocentric velocities of 8.63 km $s^{-1}$ and 13.89 km $s^{-1}$ at the time of the observation.}
\tablenotetext{c}{The observed FWHM line width of the H$_{2}$ emission lines without deconvolving the instrument profile.}
\tablenotetext{d}{The FWHM line width of the instrument profile, assuming it is Gaussian.}
\tablenotetext{e}{The FWHM resolved line width of the H$_{2}$ emission lines.}
\end{deluxetable}

\clearpage

\clearpage

\begin{deluxetable}{ccccccccccccc}
\tabletypesize{\scriptsize}
\tablecolumns{12}
\tablewidth{0pc}
\tablecaption{Ortho-to-Para ratio as a function of vibrational level \label{oprtable}}
\tablehead{
\colhead{Upper energy level [K]\tablenotemark{a}} & \colhead{Ortho H$_{2}$} &  \colhead{Para H$_{2}$}  & \colhead{OPR region A\tablenotemark{b}}  & \colhead{OPR region B}  & \colhead{OPR region C}}

\startdata

6310	&	1-0 Q(1)	&	1-0 Q(2)	&	2.15(0.09)	&	1.72(0.09)	&	1.65(0.13)	\\
6310	&	1-0 Q(1)	&	1-0 S(0)	&	2.48(0.10)	&	2.25(0.12)	&	1.97(0.15)	\\
6711	&	1-0 S(1)	&	1-0 Q(2)	&	2.15(0.06)	&	1.72(0.06)	&	1.65(0.08)	\\
6711	&	1-0 Q(3)	&	1-0 Q(2)	&	1.62(0.06)	&	1.32(0.06)	&	1.33(0.09)	\\
7268	&	1-0 S(1)	&	1-0 S(2)	&	1.31(0.07)	&	1.33(0.08)	&	1.15(0.10)	\\
7268	&	1-0 S(1)	&	1-0 Q(4)	&	1.05(0.12)	&	0.81(0.09)	&	0.47(0.08)	\\
7268	&	1-0 Q(3)	&	1-0 Q(4)	&	2.59(0.13)	&	1.80(0.10)	&	1.16(0.09)	\\
7268	&	1-0 Q(3)	&	1-0 S(2)	&	3.24(0.07)	&	2.95(0.08)	&	2.83(0.11)	\\
12322	&	2-1 S(1)	&	2-1 S(0)	&	1.56(0.15)	&	1.54(0.18)	&	1.33(0.20)	\\
12850	&	2-1 S(1)	&	2-1 S(2)	&	2.11(0.08)	&	2.19(0.11)	&	2.11(0.15)	\\
13520	&	2-1 S(3)	&	2-1 S(2)	&	2.11(0.07)	&	2.18(0.09)	&	2.11(0.12)	\\
18102	&	3-1 O(5)	&	3-2 S(2)	&	2.35(0.13)	&	2.24(0.15)	&	2.30(0.31)	\\
18736	&	3-2 S(3)	&	3-2 S(2)	&	2.35(0.10)	&	2.24(0.10)	&	2.30(0.10)	\\
22216	&	4-2 O(3)	&	4-2 O(4)	&	1.89(0.19)	&	1.66(0.16)	&	1.93(0.24)	\\
22556	&	4-2 O(5)	&	4-2 O(4)	&	1.89(0.10)	&	1.66(0.10)	&	1.93(0.16)	\\
26670	&	5-3 O(3)	&	5-3 O(2)	&	1.52(0.16)	&	1.76(0.20)	&	1.80(0.28)	\\
28188	&	5-3 Q(5)	&	5-3 Q(4)	&	1.66(0.14)	&	1.76(0.16)	&	\nodata	\\
27183	&	5-3 O(5)	&	5-3 O(4)	&	2.23(0.14)	&	2.54(0.18)	&	\nodata	\\
27626	&	5-3 O(5)	&	5-3 Q(4)	&	1.80(0.15)	&	1.54(0.16)	&	\nodata	\\
31002	&	6-4 Q(1)	&	6-4 O(2)	&	1.43(0.14)	&	1.77(0.20)	&	1.45(\nodata)	\\
31183	&	6-4 Q(1)	&	6-4 Q(2)	&	0.87(0.07)	&	1.23(0.13)	&	1.24(0.15)	\\
31482	&	6-4 Q(3)	&	6-4 Q(2)	&	0.99(0.06)	&	1.39(0.09)	&	1.23(0.12)	\\

\enddata
\tablenotetext{a}{The average value of the upper energy level of ortho and para H$_{2}$.}
\tablenotetext{b}{Uncertainties are written in parentheses.}
\end{deluxetable}

\clearpage

\clearpage

\begin{deluxetable}{cccccc}
\tabletypesize{\scriptsize}
\tablecolumns{6}
\tablewidth{0pc}\tablewidth{0pc}
\tablecaption{Gas density from comparison of the line ratios with the PDR model of \citet{Draine96} \label{pdrscom}}
\tablehead{
\colhead{Model}  & \colhead{n$_{H}$} & \colhead{UV field strength} & \colhead{$\chi^{2^{\tablenotemark{a}}}$} & \colhead{$\chi^{2^{\tablenotemark{b}}}$} & \colhead{$\chi^{2^{\tablenotemark{c}}}$}
\\   & \colhead{[cm$^{-3}$]}  & \colhead{[Habing]}  &   &  & }

\startdata

aw3o	&	10$^2$	&	1	&	77	&	23	&	29	\\
bw3d	&	10$^2$	&	10	&	91	&	26	&	31	\\
Bw3o	&	10$^3$	&	10	&	86	&	25	&	30	\\
Cw3o	&	10$^3$	&	10$^2$	&	52	&	16	&	26	\\
Gw3o	&	10$^4$	&	10$^2$	&	25	&	15	&	26	\\
Hw3o	&	10$^4$	&	10$^3$	&	26	&	17	&	27	\\
Lw3o	&	10$^5$	&	10$^3$	&	32	&	38	&	39	\\
Mw3o	&	10$^5$	&	10$^4$	&	57	&	53	&	48	\\
Qw3o	&	10$^6$	&	10$^4$	&	75	&	85	&	63	\\

\enddata
\tablenotetext{a}{Chi-square $\chi^{2}$ of region A.}
\tablenotetext{b}{Chi-square $\chi^{2}$ of region B.}
\tablenotetext{c}{Chi-square $\chi^{2}$ of region C.}

\end{deluxetable}

\clearpage



\clearpage

\clearpage

\clearpage

\clearpage

\clearpage

\clearpage

\clearpage

\clearpage

\clearpage

\clearpage

\clearpage

\clearpage

\clearpage

\clearpage

\clearpage

\clearpage

\clearpage

\clearpage

\clearpage

\end{document}